\documentclass[twocolumn,pre,floatfix]{revtex4-2}
\usepackage{psfrag,epsfig,amsfonts,amssymb,amsmath,wasysym,bm}
\usepackage{dcolumn}
\usepackage{bbold}
\usepackage[normalem]{ulem}
\usepackage{color}
\usepackage{tabularx}
\usepackage{tikz}
\usepackage{ulem}

\usepackage[inline]{enumitem} 

\usepackage{hyperref}

\newcommand{\G}{\Lambda}
\newcommand{\At}{\langle A\rangle_{\! t}}

\newcommand{\dof}{\kappa}
\newcommand{\gap}{\gamma^{0}}
\newcommand{\Att}{{\cal A}_{t}}
\newcommand{\TT}{{\cal T}}
\newcommand{\pmax}{p_{\rm max}}

\newcommand{\amax}{a_{\rm max}}
\newcommand{\amin}{a_{\rm min}}
\newcommand{\Da}{\Delta_{\!A}}

\newcommand{\CC}{{\mathbb C}}

\newcommand{\tr}{\mbox{Tr}}
\newcommand{\hr}{{\cal H}}
\newcommand{\ord}{{\cal O}}

\providecommand{\norm}[1]{\|#1\|}

\newcommand{\lmat}{\left( \begin{matrix}}	
\newcommand{\rmat}{\end{matrix} \right)}	
\newcommand{\figref}[1]{Fig.~\protect\ref{#1}}

\newcommand{\xref}[1]{\protect\ref{#1}}

\begin{document}

\title{Non-equilibration, synchronization, and time crystals in isotropic Heisenberg models}

\author{Peter Reimann}
\author{Patrick Vorndamme}
\author{J\"urgen Schnack}
\affiliation{Faculty of Physics, 
Bielefeld University, 
33615 Bielefeld, Germany}
\date{\today}

\begin{abstract}
Isotropic but otherwise largely arbitrary 
Heisenberg models in the presence of a homogeneous magnetic field
are considered, including various integrable, non-integrable, as 
well as disordered examples, and not necessarily restricted to one 
dimension 
or
short-range interactions.
Taking for granted that the non-equilibrium initial condition 
and the spectrum of the field-free model satisfy some very weak requirements,
expectation values of generic observables are analytically shown to
exhibit permanent long-time oscillations, thus ruling out equilibration.
If the model (but not necessarily the initial condition) is translationally invariant, 
the long-time
oscillations are moreover shown to 
exhibit synchronization in the long run, meaning that they are invariant under arbitrary translations of the observable.
Analogous long-time oscillations are also recovered for 
temporal correlation functions when the system is 
already at thermal
equilibrium from the outset, thus realizing a so-called time crystal.
\end{abstract}

\maketitle

\section{Introduction}
\label{s1}

A macroscopic system without 
external perturbations
approaches
a steady equilibrium 
state 
after sufficiently long times,
no matter how far from equilibrium it started out. 
On the phenomenological level, this is an extremely 
well-established fact both in everyday life and under 
controlled laboratory conditions.
More precisely speaking, in every single run of an experiment,
one may still encounter certain statistical or quantum mechanical 
fluctuations, especially for microscopic observables, but 
on the average over many repetitions of the experiment, 
the expectation value will
closely approach some constant equilibrium value 
in the long run.
On the other hand, a satisfactory theoretical understanding 
of these 
empirical observations
in terms of the underlying fundamental 
laws of quantum mechanics still remains a challenging 
open question, both qualitatively and quantitatively,
to which a considerable amount of experimental, numerical,
and analytical efforts have been devoted in recent years 
\cite{mor18,dal16,gog16,lan16,ued20,nan15}.

Obviously, a particularly fascinating 
endeavor in this context is 
to identify cases which 
give rise to
certain deviations from the 
above-mentioned standard 
scenario.
For instance,
it has been discovered that models exhibiting integrability or 
many-body localization may 
not entail thermalization, meaning that
expectation values do not closely approach
the pertinent
canonical or microcanonical values
predicted by 
equilibrium statistical mechanics after sufficiently long times
\cite{mor18,dal16,gog16,lan16,ued20,nan15}.
Nevertheless, 
generically they
still exhibit equilibration,
meaning that
the time-dependent expectation values 
stay extremely close to a constant value for the vast majority
of all sufficiently late times, i.e., apart from
the transient relaxation processes during some initial 
time-interval, and apart from the well-known, exceedingly 
rare but unavoidable quantum recurrence or revival effects
\cite{rei08,lin09,sho11,sho12,rei12,bal16,rid23}.

At the focus of our present work are many-body
systems 
whose expectation values
do not even equilibrate in the above sense, but rather exhibit
permanent 
long-time oscillations.
Leaving aside trivial cases like non-interacting
systems or perfect harmonic oscillators, 
related previously proposed examples 
that may come to one's mind are the  ``quantum Newton's cradle'' 
experiment by Kinoshita, Wenger, and Weiss \cite{kin06},
the exploration of Rydberg-atom quantum simulators 
by Bernien et al. \cite{ber18}, or
the numerical study by Banuls, Cirac, and Hastings~\cite{ban11}.
However, it was later discovered that in fact
all those examples
ultimately still must exhibit equilibration when monitoring
the dynamical evolution over sufficiently long times
\cite{li20,mbs,kim15,lin17,far17}.
On the other hand, analytically provable absence of equilibration 
in the context of many-body quantum scars
has been recently 
established
for various abstract models
in combination with special initial conditions 
\cite{mbs}, yet their significance with regard to real-world 
systems still remains to be explored.

In our present work, we 
focus on
one of the simplest and best-established
many-body quantum systems, 
namely the isotropic Heisenberg model
with a homogeneous magnetic field.
Besides the original and most common version of the model, 
also various generalizations and modifications 
will be covered, including non-intergable 
systems 
and disorder in the form of randomized interactions.
The only indispensable prerequisites are that the field-free model 
must be SU(2) symmetric, i.e.\ isotropic,
the external field must be spatially homogeneous,
and
the energy levels must
satisfy some rather weak and generic assumptions.

Our first main objective is to analytically demonstrate and 
numerically illustrate the typical 
occurrence
of 
non-equilibration in the form of everlasting oscillations 
in such systems.
In particular,
this behavior is restricted neither to 
special initial conditions nor to integrable models.

Furthermore, 
we analytically show that those oscillations entail synchronization 
under the additional condition that the model -- but not 
necessarily the initial condition -- is translationally invariant.

Turning to systems at thermal equilibrium, we finally establish
the generic 
occurrence
of analogous long-time oscillations for 
dynamic (time-dependent) correlation functions, and we discuss their implications 
with respect to the topic of time crystals
\cite{zal23,han22,ven19,med20,wat15,wat20,hua19}.

In terms of these main findings, but also methodologically,
our present paper is closely related in a variety of
different respects to a considerable number of previous
works, 
including Refs.~\cite{med20,vor21,rei08,lin09,sho11,sho12,rei12,bal16,alh20,wat15,wat20,hua19}.
Since an adequate comparison is only possible
on the basis of a minimal amount of formal definitions,
such a more detailed discussion of pertinent previous 
works will be provided at various places throughout the 
paper.


\section{General framework}
\label{s2}

We consider a Heisenberg model on an arbitrary 
(not necessarily one-dimensional) lattice, 
whose sites are labeled by $i$. We denote by $\G $ 
the set of all possible lattice sites, and by $\dof$ their 
total number.
Alternatively, $\dof$ may thus be viewed as the
system size or as the number of degrees of freedom.
The single-site spin operators are indicated by
vectors $\vec s_i$ with three components 
$s_i^{a}$, $a\in\{x,y,z\}$, while the single-site
spin quantum number is given by the same
integer or half-integer $s$ on every site.

Denoting the components of the total spin by 
\begin{eqnarray}
S^a:=\sum_{i\in \G }  s_i^a
\ ,
\label{1}
\end{eqnarray}
the considered Hamiltonians must be of the general form
\begin{eqnarray}
H & := & H_0 + h\, S^z \, ,
\label{2}
\\
H_0 & := & \sum_{i,j\in \G } J_{ij} \, \vec s_i\cdot\vec s_j \ ,
\label{3}
\end{eqnarray}
where the magnetic field $h$ and the coupling constants 
$J_{ij}$ are, for the time being, still largely arbitrary model parameters.

Since the Hamiltonian $H_0$ in (\ref{3}) is spatially 
isotropic, it possesses SU(2) symmetry 
and thus commutes with $S^a$ for all $a\in\{x,y,z\}$.
As a consequence, the eigenvectors of $H_0$ can be chosen 
so that they are simultaneously eigenvectors of $S^z$
as well as of  $\vec S^2:=(S^x)^2+(S^y)^2+(S^z)^2$,
and thus can be written as $|n,l\rangle$ 
with the properties
\begin{eqnarray}
H_0 |n, l\rangle & = & E_n^0 \, |n, l\rangle
\ ,
\label{4}
\\
S^z |n,l\rangle & = & l \, |n,l\rangle
\ ,
\label{5}
\\
\vec S^2 |n,l\rangle & = & L_n(L_n+1)\, |n,l\rangle
\ .
\label{6}
\end{eqnarray}
Here, the indices $n\in\{1,...,N\}$ label the energy eigenvalues,
the $l \in \{-L_n,...,L_n\}$ are the total magnetic quantum 
numbers, 
while the $L_n$ are 
positive integers or half-integers, often denoted as 
total spin quantum numbers.
In other words, for any given $n$, the energies
$E_n^0$ are $(2L_n\!+\!1)$-fold degenerate 
with
spin multiplets $\{|n,l\rangle\}_{l=-L_n}^{L_n}$.
Traditionally, those simultaneous eigenvectors of 
$H_0$, $\vec S^2$, and $S^z$ are often 
denoted as $|n,L_n,l\rangle$, but since the
$L_n$'s are unique functions of the $n$'s,
we employ the shorter notation $|n,l\rangle$.
One readily verifies that $0\leq L_n \leq \dof s$, 
and one can evaluate how many 
eigenvectors
belong to a certain $l$ or $L_n$ \cite{BSS:JMMM00}, 
but for the rest, the actual quantitative
value of $L_n$ belonging to any given 
$n$ (or $E_n^0$) is
in general quite difficult to tell; see also Appendix \ref{app1}.
We finally remark that the energies $E_n^0$ are 
generically expected to be pairwise different, 
but that this property is not
actually required in most of our subsequent
explorations.

Exploiting (\ref{2}), (\ref{4}), (\ref{5}) it follows that
\begin{eqnarray}
H\, |n,l\rangle & = & E_{nl} \, |n,l\rangle
\ ,
\label{7}
\\
E_{nl} & := & E_n^0+l\, h
\ .
\label{8}
\end{eqnarray}
The eigenvectors $|n,l\rangle$ are thus independent
of $h$, while the above-mentioned degeneracies of 
the eigenvalues for $h=0$ are expected to be 
generically lifted for $h\not =0$
(Zeeman splitting).

Given any pure or mixed initial state $\rho(0)$,
its time evolution is governed by the 
von Neumann equation, resulting at time $t$ in the state
$\rho(t)=e^{-iHt}\rho(0)e^{iHt}$ ($\hbar=1$).
Accordingly, the expectation value of any observable 
(Hermitian operator) $A$ at time $t$ is given by
\begin{eqnarray}
\At:=\tr\{\rho(t) A\} \ .
\label{8a}
\end{eqnarray}
By employing the eigenvalues and eigenvectors of $H$ 
from (\ref{7}) and (\ref{8}) it 
follows that
\begin{eqnarray}
\At = \sum_{mnkl} \rho_{mn}^{k,l} A_{nm}^{l,k}\, e^{i(E_n^0-E_m^0+[l-k]h)t}
\ ,
\label{9}
\end{eqnarray}
where the sum is tacitly restricted to indices $m,n,k,l$ within their 
admitted range as specified below (\ref{6}), and
where the matrix elements $\rho_{mn}^{k,l}$ and $A_{nm}^{l,k}$ are defined as
\begin{eqnarray}
\rho_{mn}^{k,l} & := & \langle m,k|\rho(0)|n,l\rangle
\ , 
\label{10}
\\
A_{nm}^{l,k} & := & \langle n,l | A|m,k\rangle
\  .
\label{11}
\end{eqnarray}
Going over from the summation index $l$ in (\ref{9}) to $\nu:=l-k$
then yields
\begin{eqnarray}
\At & = & \sum_\nu f_{\nu}(t) \, e^{i\nu h t}
\ ,
\label{12}
\\
f_{\nu}(t) & := & \sum_{mn} e^{i(E_n^0-E_m^0)t}\sum_k  \rho_{mn}^{k,k+\nu} A_{nm}^{k+\nu,k}
\ .
\label{13}
\end{eqnarray}

One readily verifies that $f_{-\nu}(t)=f^\ast_\nu(t)$, hence (\ref{12}) 
could also be rewritten as a purely real Fourier series.
Since the eigenvectors $|n,l\rangle$ in (\ref{4})
and thus in (\ref{7}) are independent of $h$, the same property is inherited by the
matrix elements in (\ref{10}) and (\ref{11}), and finally by the functions 
$f_\nu(t)$ in (\ref{13}).
In other words, the only $h$-dependence in (\ref{12}) arises via the exponential 
factors on the right-hand side.

%
\subsection{Model classification}
\label{s21}

The general structure in (\ref{1})-(\ref{3}) 
still covers a wide variety of models in one or more 
dimensions, whose interactions may 
be of short- or long-range character, 
and may even exhibit various kinds of disorder
with concomitant many-body localization effects \cite{nan15}.
Moreover, also our assumption that all lattice sites exhibit the same 
spin quantum number $s$
can be readily relaxed.

We emphasize that these models (\ref{1})-(\ref{3})
include many examples which are commonly 
considered 
as being either integrable or 
non-integrable,
even though the precise meaning of ``integrability'' 
is still not entirely clear \cite{dal16,gog16}.
Independently of such still unsettled subtleties,
for our present purposes it seems reasonable to require
that whether a given model in (\ref{1})-(\ref{3}) is considered 
as (non-)integrable
should {\em not} depend on the value of 
the external field $h$.
The reason is that since the eigenvectors in (\ref{7}) 
are independent of $h$, and the dependence of 
the eigenvalues in (\ref{8}) on $h$ is rather trivial,
it would not be satisfying if a transition from integrable 
to non-integrable would be achievable by simply changing the 
value of $h$.

%
\section{Main results}
\label{s3}

Our first main result consists in the prediction that,
for sufficiently large systems,
the expectation values in (\ref{12}) 
can be approximated very well by
\begin{eqnarray}
\Att :=\sum_{\nu} \bar f_{\nu} \, e^{i\nu ht}
\label{14}
\end{eqnarray}
for the vast majority of all sufficiently late times $t$,
where $\bar f_{\nu}$ essentially amounts to the long-time 
average of $f_\nu(t)$ from (\ref{13}).
More precisely speaking,
\begin{eqnarray}
\bar f_{\nu}:= {\sum_{mnk}}' \rho_{mn}^{k,k+\nu} A_{nm}^{k+\nu,k}
\ ,
\label{15}
\end{eqnarray}
where the prime symbol indicates that the summation is
restricted to indices $m$ and $n$ with the property $E_m^0=E_n^0$.
In the generic case that all energies $E_n^0$ are pairwise different
(see below Eq.~(\ref{6})), this boils down to
\begin{eqnarray}
\bar f_{\nu} = \sum_{nk} \rho_{nn}^{k,k+\nu} A_{nn}^{k+\nu,k}
\ .
\label{16}
\end{eqnarray}
More generally, the same simplification (\ref{16}) of (\ref{15})
also applies to cases where either $\rho_{mn}^{k,l}$ or 
$A_{nm}^{l,k}$ vanishes whenever $m\not=n$ and 
$E_m^0=E_n^0$ (degeneracies).
We also recall that similar restrictions as below (\ref{9}) are 
understood to apply to the sums in (\ref{15}) and (\ref{16}).

Before providing the quantitative analytical details of the above prediction,
let us mention a non-rigorous argument of 
how the emergence of such a result may be 
intuitively understood:

Indicating the average over all times $t\geq 0$ 
by $\langle \,\cdot\,\rangle_{\!\infty}$, 
we can conclude that
$\langle e^{i(E_n^0-E_m^0)t} \rangle_{\!\infty}$ 
equals unity if $E_m^0=E_n^0$ and zero otherwise.
Together with (\ref{13}) and (\ref{15}) it follows that
$\langle f_{\nu}(t)\rangle_{\!\infty} = \bar f_\nu$.
Moreover, for sufficiently large systems, the number of summands 
on the right-hand side of (\ref{13}) may be expected to become
very large. 
Incidentally, in view of the restrictions mentioned 
below Eq.~(\ref{9}), a more rigorous justification of this 
argument for any single $\nu$ may be difficult.
Taking it for granted nevertheless,
the key point now consists in the heuristic conjecture 
that this large number of summands in (\ref{13})
entails some kind of 
``dephasing effect'',
with the result that all the summands with 
$E_m^0\not=E_n^0$
effectively cancel each other 
in sufficiently good approximation.
As a consequence,
every $f_\nu(t)$ in (\ref{13}) is conjectured to stay near its time 
average (\ref{15}), and  hence the expectation values (\ref{12})
to stay near $\Att$ from (\ref{14}).

Next we turn to a more rigorous foundation of 
our prediction.
In doing so, we proceed in three steps.
First, the two most important quantities appearing 
in our main analytical result are introduced.
Next, the analytical result itself is presented
and discussed.
Finally, the actual derivation of the result is
provided in Appendix \ref{app1}.

For an instructive numerical illustration of those
general predictions, we refer to Sec.~\ref{s35}.

%
\subsection{Level populations and energy gaps}
\label{s31}

According to the first remark below Eq.~(\ref{8}), 
the quantity $\langle n,l |\rho(0)| n,l\rangle$
is independent of the magnetic field $h$.
Moreover, it can be identified with the population of the energy 
eigenstate $|n,l\rangle$ by the initial state $\rho(0)$.
Likewise,
\begin{eqnarray}
\pmax & :=& \max_{n,l} \langle n,l |\rho(0)| n,l\rangle
\label{17}
\end{eqnarray}
thus amounts to  the {\em maximal level population}
and is $h$-independent.
We also note that the corresponding time-dependent level 
populations $\langle n,l |\rho(t)| n,l\rangle$ are actually independent of $t$, 
as can be seen by rewriting them in the form (\ref{8a})
with $A:=|n,l\rangle\langle n,l|$ and then exploiting (\ref{9}).

Next we focus on an arbitrary but fixed pair of indices 
$(m,n)$ with 
the property $E_m^0\not =E_n^0$,
and we count all possible index pairs
$(m',n')$ whose energy gaps $E_{m'}^0-E_{n'}^0$
are equal to the given reference gap $E_m^0-E_n^0$.
The number of those pairs $(m',n')$
is denoted as
$\gap_{mn}$.
For obvious reasons, this number $\gap_{mn}$ is called
the degeneracy of the energy gap $E_m^0-E_n^0$,
and it has the properties that $\gap_{mn}\geq 1$ and
$\gap_{nm}=\gap_{mn}$.
Specifically, if $\gap_{mn}=1$ then $E_m^0-E_n^0$
is called a non-degenerate energy gap.
Finally, the {\em maximal energy gap degeneracy}
is defined as
\begin{eqnarray}
\gap:=\max_{m,n} \gap_{mn}
\ ,
\label{18}
\end{eqnarray}
where the maximum is taken over all pairs
$(m,n)$ with non-vanishing energy gaps $E_{m}^0-E_{n}^0$ 
\cite{sho12}.

We close with two side remarks: 
(i) 
The above defined quantities $\gap_{mn}$ and $\gap$ refer, 
as indicated by the superscript ``$0$'', to the unperturbed 
system, and as such are independent of $h$ for trivial reasons.
(ii) As already mentioned below (\ref{6}), we do not require 
that all $E_n^0$ are pairwise different,
with the following implication 
with regard to $\gap$:
Denoting for any given $n$ the number
of indices $k$ with the property
$E_k^0=E_n^0$ by $\mu(n)$ (``multiplicity of $E_n^0$'')
it readily follows that $\gap_{mn}\geq \mu(m)\mu(n)$,
and hence that $\gap\geq \mu_{\max}^2$,
where $\mu_{\max}:=\max_n\mu(n)$ is 
the maximal number of pairwise identical
energies $E_n^0$. 
Even for integrable systems such as spin-1/2 rings 
the maximal number of pairwise identical energies 
increases only by a factor of order 2 and only in 
certain Hilbert-subspaces \cite{YAS:JPA02}.
On the other hand, even if all $E_n^0$ are pairwise different
and thus $\mu_{\max}=1$, it is still possible that $\gap>1$.

%
\subsection{Main analytical prediction}
\label{s32}

Employing the definitions (\ref{17}) and (\ref{18}),
and indicating the temporal average over an interval $[0,T]$ 
by the symbol $\left\langle \,\cdot\,\right\rangle_{T}$, 
it is shown in Appendix \ref{app1} that the
mean square deviation of the ``true'' expectation 
values (\ref{12}) from the auxiliary function (\ref{14})
obeys for all sufficiently large $T$ the inequality
\begin{eqnarray}
\left\langle [\At - \Att ]^2 \right\rangle_{T} 
& \leq & \gap\, (2s \dof\!+\!1)^2\, \Da^2\, \pmax
\ ,
\label{19}
\end{eqnarray}
where $s$ is the single-spin quantum number and $\dof$ the system 
size 
(see above Eq.~(\ref{1})).
Furthermore, $\Da$ is the measurement range of the observable $A$,
i.e., the difference between the largest and smallest possible 
measurement outcomes, or equivalently, eigenvalues of $A$.

Our first remark is that the right-hand side of (\ref{19}) is 
independent of the magnetic field $h$ in (\ref{2}).

Our second remark is that 
the level density of a many-body system is commonly 
known or expected \cite{mor18,dal16,tas18}
to grow exponentially fast with the system size $\dof$.
Hence, the level density will become 
extremely high for macroscopically large systems,
and it will be practically impossible in a real 
experiment to notably populate
only a small number of 
eigenstates $|n,l\rangle$.
Rather, one expects 
\cite{rei08,rei12,bal16}
that the number of  non-negligibly 
populated 
levels 
will still be exponentially large in $\dof$. 
Recalling Eq.~(\ref{17}), 
and that the sum of all level 
populations must be unity, 
one thus expects \cite{rei08,rei12,bal16}
that a very rough order of magnitude estimate of the form
\begin{eqnarray}
\pmax \approx \exp\{{-\ord(\dof)}\}
\label{20}
\end{eqnarray}
will be generically fulfilled under all experimentally 
realistic circumstances.
For some particularly important examples,
a more detailed confirmation of this property
will be provided in Sec.~\ref{s33} below.

Our third remark is that, obviously, no significant conclusion 
about the expectation values in (\ref{9}) can be drawn 
without any knowledge whatsoever regarding the energies 
$E_n^0$ appearing on the right-hand side.
On the other hand, these energies are in general not explicitly 
known in sufficient quantitative detail.
An exception is given by models that are analytically solvable by 
means of the Bethe ansatz, but in practice this is of little use
for our present purposes.
For instance, already one of the simplest and most important 
features of the energies $E_n^0$,
namely the so-called level statistics (probability 
distribution of the distances between neighboring energy 
levels), is not analytically available for practically any
quantum many-body system of physical interest, 
including our present Heisenberg 
models of the general form (\ref{2}).
However,
it is commonly taken for granted 
 -- based on heuristic arguments and ample numerical evidence --
that the level statistics tends to some well-defined 
and reasonably smooth asymptotics in the thermodynamic limit.
Moreover, this asymptotics is often expected to be 
close to, for instance, a Wigner-Dyson or a Poisson
distribution, but such ``details''  do not matter here.

Our present assumption regarding the 
energies $E_n^0$ is 
in essence
quite similar in spirit 
to these
common assumptions regarding the level statistics.
Namely, we assume that the maximal energy gap
degeneracy in (\ref{18}) grows at most subexponentially 
with the system size $\dof$.
Indeed, this is 
closely related to requiring that
the level statistics does not develop delta-peaks
in the thermodynamic limit.
In particular, this also means that the maximal 
number of pairwise identical energies
$E_n^0$ must grow at most 
subexponentially with $\dof$, see 
remark (ii) at the end of Sec.~\ref{s31}.

Finally, it is also noteworthy that our above assumptions 
regarding $\pmax$ and $\gap$ are by now very well-established 
in the context of equilibration 
in many-body 
quantum systems, and that there 
exists
essentially no rigorous 
analytical result in this context which is valid without taking for granted 
the same or some very similar assumptions
\cite{mor18,gog16,rei08,lin09,sho11,sho12,rei12,bal16,far17,tas98,sre99,mul15,imb16,gal18,wil19}.

Altogether, we thus can and will take our above assumptions
regarding $\pmax$ and $\gap$ for granted.
For large $\dof$, the small factor $\pmax$ in (\ref{20}) then overrules
by far the factors $\gap$ and $\dof^2$ on the right-hand side of 
(\ref{19}), implying that the time-averaged variance  
on the left-hand side of (\ref{19}) will be exponentially 
small compared to the (squared) measurement range 
$\Da$ of the observable.
In turn, this is only possible if the difference
$\At - \Att$ is unmeasurably small
(below the resolution limit of the measurement device $A$)
for the overwhelming majority of all time points $t\in[0,T]$.
As already said in the Introduction,
time points $t$ belonging to the complementary, exceedingly
small minority are generically expected to occur during the
initial transient relaxations processes, and on the occasion
of the well-known, exceedingly rare, but unavoidable 
quantum recurrences or revivals,
see, e.g., Ref. \cite{rid23} and further references therein.
The initial relaxation may in fact be viewed as one of them.
Moreover, the origin of those revivals is closely related to
the fact that the sum in (\ref{9}) is a quasi-periodic function of
$t$.
All these complications are effectively taken into account by 
our requirement above (\ref{19}) that $T$ must be sufficiently
large.

In summary, our main finding is that the deviations between $\At$
and $\Att$ will be negligibly small for the overwhelming majority
of all sufficiently late times $t$, symbolically indicated as
\begin{eqnarray}
\At \rightsquigarrow \Att
\ .
\label{21}
\end{eqnarray}

Incidentally,
similar methods as in the derivation of our present result in
Appendix \ref{app1}
have been previously adopted, e.g., in Ref.~\cite{rei08,lin09,sho11,sho12,rei12,bal16}
in the context of {\em equilibration}, i.e., for the purpose
to show that the expectation 
values $\At$ remain -- under suitable conditions on the 
Hamiltonian $H$, the initial state $\rho(0)$, and the observable $A$ --
very close to some {\em constant} value for the vast 
majority of all sufficiently late times $t$.
Obviously, such a prediction of equilibration
cannot apply to our present models (\ref{2}) with $h\not =0$ 
since they generically give rise to everlasting oscillations 
of $\At$, see also Sec.~\ref{s34} below.
The main reason is that the energies $E_{nl}$ in (\ref{8})
violate (for $h\not=0$) the corresponding requirements
in Refs.~\cite{rei08,lin09,sho11,sho12,rei12,bal16}
regarding the maximally admissible degeneracy
of the pertinent energy gaps.
Indeed, one finds that our present models entail some
exponentially large sets of degenerate energy gaps:
For instance, considering two arbitrary but fixed indices $l$ and $l'$
we can conclude from Eq.~(\ref{8}) that the energy gaps $E_{nl}-E_{nl'}$ 
are equal for all possible values of $n$, while the total 
number of all those $n$ values is often 
expected to be exponential in the system size.
Likewise, for any given set of indices $n,l,n',l'$ the energy
gaps $E_{n(l+l'')}-E_{n'(l'+l'')}$ are equal for all possible
values of $l''$.
As a consequence, for $h\neq 0$ our models violate one of the central 
preconditions for equilibration established in Refs.~\cite{rei08,lin09,sho11,sho12,rei12,bal16}.

In contrast, the maximal degeneracy of energy gaps employed in
(\ref{18}) is a property of the {\em unperturbed} ($h=0$)
energies $E_n^0$, {\em not} of the energies
$E_{nl}$ pertaining to
the actually considered model Hamiltonian $H$ in (\ref{2})
with $h\not=0$.
In passing, we also remark that, according to Refs.~\cite{rei08,lin09,sho11,sho12,rei12,bal16}, 
it is the degeneracy of these gaps 
for $h\not = 0$
which prohibits equilibration, 
{\em not} their commensurability, as 
speculated, e.g., in \cite{boo20}.

\subsection{Canonical quenches}
\label{s33}

In view of (\ref{17}) we can conclude that
$(\pmax)^2$ is upper bound by 
$\sum_{nl}  \langle n,l |\rho(0)| n,l\rangle^2$ and hence by 
$\sum_{nlmk}  |\langle n,l |\rho(0)| m,k\rangle|^2= \tr\{[\rho(0)]^2\}$,
implying
\begin{eqnarray}
\pmax \leq \sqrt{ \tr\{[\rho(0)]^2\} }
\ .
\label{22}
\end{eqnarray}

As a particularly simple and interesting example, 
let us assume that the initial state is given by a thermal 
Gibbs state (canonical ensemble) of the form 
\begin{eqnarray}
\rho(0) = \tilde Z^{-1} e^{-\beta \tilde H}\ ,\ \ \tilde Z:=\tr\{ e^{-\beta \tilde H} \}
\ ,
\label{23}
\end{eqnarray}
where $\tilde H$ is in general different from the Hamiltonian 
$H$ in (\ref{2}) which governs the subsequent 
temporal evolution of $\rho(0)$.

For instance, one may choose $\tilde H$ to be of the general form
\begin{eqnarray}
\tilde H & := & H_0 + \sum_{i\in \G } \vec h_i\cdot \vec s_i
\ ,
\label{24}
\end{eqnarray}
thus differing from $H$ in (\ref{2}) with respect to the
direction and possibly also the magnitude of the externally 
applied magnetic field at any of the lattice sites $i$.
Further examples of how to choose physically reasonable 
$\tilde H$'s are rather obvious,
see also Sec.~\ref{s35} below.

From a different viewpoint, the system may thus
be considered as being
at thermal equilibrium for $t<0$ and experiencing an instantaneous
``quantum quench'' at $t=0$, with pre-quench 
Hamiltonian $\tilde H$ and post-quench Hamiltonian $H$.

Exploiting that the
free energy $F_\beta$
associated with the canonical ensemble (\ref{23}) 
obeys the relation $e^{-\beta F_\beta}=\tr\{e^{-\beta \tilde H}\}$,
one can conclude that $\tr\{[\rho(0)]^2\}=e^{-2\beta G_{\! \beta}}$
with $G{\! _\beta}:=F_{2\beta}-F_{\beta}$.
Taking for granted that the pre-quench system exhibits
generic thermodynamic properties, it follows that
$G{\! _\beta}$ is an extensive quantity.
Hence, $\tr\{[\rho(0)]^2\}$ decreases exponentially 
with the system size $\dof$, and likewise for $\pmax$
in (\ref{22}).

Altogether, we thus have rigorously verified (\ref{20})
for initial conditions of the  canonical form (\ref{23}).
The same conclusion can also be readily recovered
for microcanonical instead of canonical initial states
$\rho(0)$.

\subsection{Permanent oscillations}
\label{s34}

To begin with, we note that
$\bar f_\nu$ in (\ref{15}) must be zero if $|\nu|> 2\dof s$
as a consequence of the restrictions on the summation indices below (\ref{9})
(the detailed reasoning is worked out 
below Eq.~(\ref{a14})).
Furthermore, one can infer from (\ref{10}), (\ref{11}),
and (\ref{15}) that $\bar f_{-\nu}=\bar f_{\nu}^\ast$.
Representing the complex numbers $\bar f_\nu$ 
in the polar form $|\bar f_\nu|e^{i\varphi_\nu}$, 
we thus can rewrite (\ref{14}) as
\begin{eqnarray}
\Att =\bar f_0+ 2 \sum_{\nu=1}^{2\dof s} |\bar f_{\nu}| \, \cos(\nu ht+\varphi_\nu)
\ .
\label{25}
\end{eqnarray}

Generically, the quantities $\bar f_\nu$ in (\ref{15}) are 
not expected to identically vanish for all 
$\nu\not=0$,
hence (\ref{25}) together with (\ref{21}) implies 
the occurrence of permanent oscillations for all
sufficiently late times $t$.
Exceptional cases, tailored such 
that $\bar{f}_\nu=0$ for all $\nu\not=0$, 
will be addressed later in Sec. \ref{s6}.

As an aside, we 
remark that
the quantity $\bar f_0$ in (\ref{25}) obviously represents the long-time average of
$\Att$. In the generic case that all energies $E_n^0$ are pairwise different
(see also below Eq.~(\ref{6})), $\bar f_0$
can be further rewritten by means of 
(\ref{16}) and the so-called 
{\em diagonal ensemble}
\begin{eqnarray}
\rho_{\rm{dia}}:=\sum_{nl} \rho_{nn}^{l,l} \, |n,l\rangle\langle n,l |
\label{26}
\end{eqnarray}
in the form
\begin{eqnarray}
\bar f_0 = \tr\{\rho_{\rm{dia}} A\}
\ .
\label{27}
\end{eqnarray}

We also remark that our present oscillatory long-time effects are 
similar to those recently 
discovered in the ground-breaking work \cite{med20}. 
A first important difference is that Ref.~\cite{med20} is mainly focused on the
one-dimensional spin-1/2 XXZ-model (which is integrable), while our present model
class also covers, for instance, various non-integrable and disordered systems
(cf. Sec.~\ref{s21}).
The second important difference is that 
the findings reported in Ref.~\cite{med20} are
mainly based on non-rigorous arguments and numerical 
evidence,
adopting 
some rather special initial states and observables.
Finally, the 
prediction of permanent oscillations in Ref.~\cite{med20}
only applies to a quite restricted (discrete) subset of the XXZ spin chain's 
anisotropy parameter values.

\subsection{Numerical examples}
\label{s35}

The subsequent numerical examples are chosen to illustrate 
our two main analytical findings for Hamiltonians of the
general form (\ref{1})-(\ref{3}): 
(I) permanent oscillations, and (II) 
synchronization of these oscillations 
in case of translationally invariant Hamiltonians,
see also Sec.~\xref{s4} below.
To this end, we numerically explore the behavior 
of
the following 
four specific models:
(i) A spin ring (i.e.\ periodic boundary conditions)
with unperturbed Hamiltonian
\begin{eqnarray}
H_0
& := & \sum_{i = 1}^{\dof} J_i \vec s_i\cdot \vec s_{i+1}
\ ,
\label{H0R}
\end{eqnarray}
exhibiting quenched disorder by choosing the interactions 
$J_i$ as independent, identically distributed random numbers.
(ii) The same spin ring model as in (\ref{H0R}), but now with
identical couplings $J_i$ for all $i$ (no disorder).
(iii) A two-dimensional (2D) 
$5\times 5$ square lattice model
with identical nearest-neighbor interactions,
open boundary conditions in both directions, 
and unperturbed Hamiltonian
\begin{eqnarray}
\label{H02D}
H_0
& := & J \sum_{<i,j>} \vec s_i\cdot \vec s_{j}
\ . 
\end{eqnarray}
(iv) The same square lattice model as in (\ref{H02D}), 
but now with periodic boundary conditions in both 
directions.
Similar to (\ref{1})-(\ref{3}),
an additional homogeneous magnetic field in $z$-direction 
is applied during time evolution 
in all four cases (i)-(iv).
Accordingly, the spin ring without disorder represents an 
integrable model, whereas all the other examples (i), (iii), and (iv)
are commonly considered as non-integrable, see also Sec.~\ref{s21}.
Moreover, (ii) and (iv) are so-called translationally
invariant models (see also Sec. \ref{s4} below),
while (i) and (iii) are not.

\begin{figure}
\includegraphics[scale=0.20]{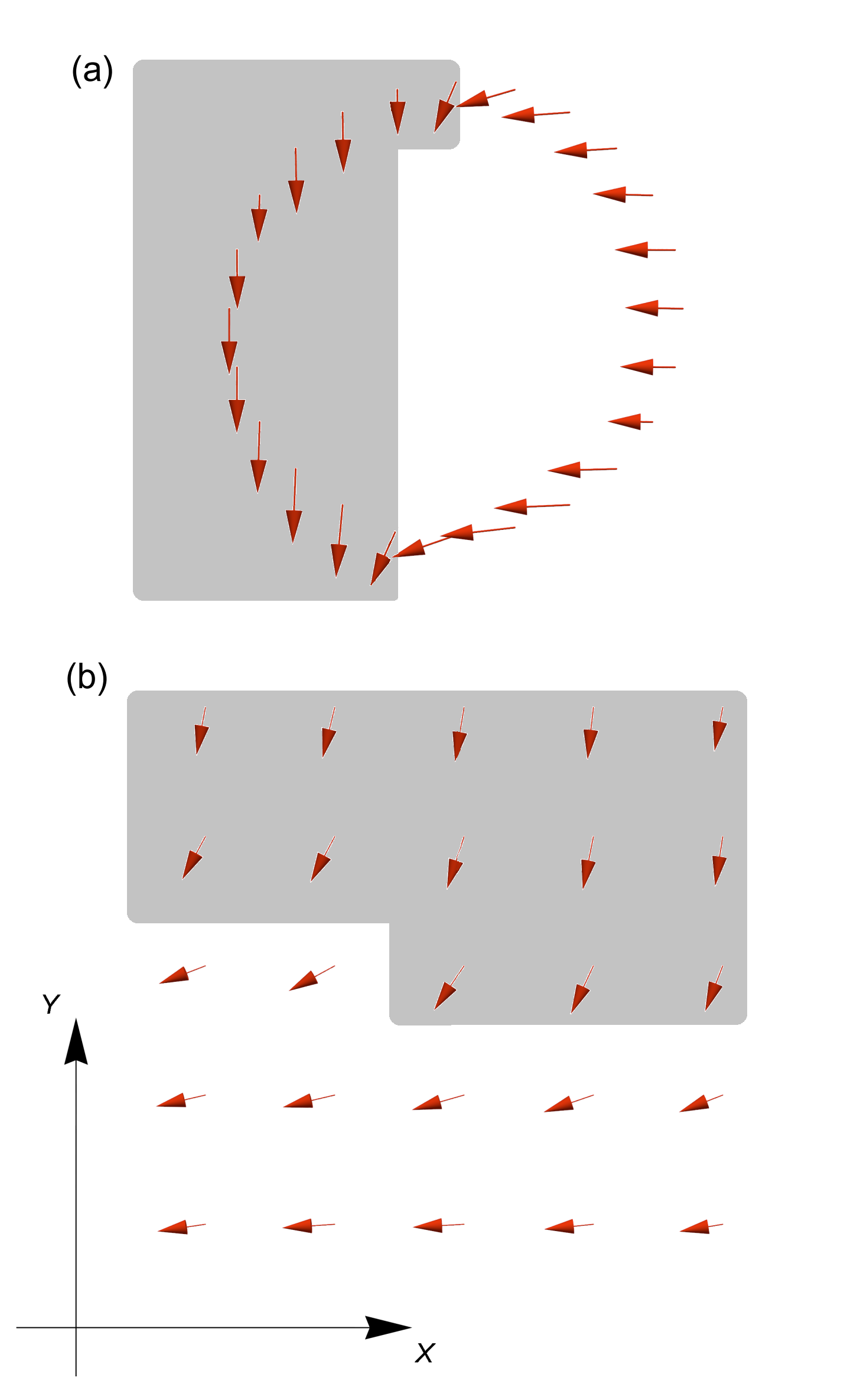}
\caption{
Red arrows: Visualization of the projections to the $(x,y)$-plane 
of the expectation values of the local spin vector operators 
$\vec s_i$ with respect to the initial state \eqref{psi0}.
(a): Spin ring models (i) and (ii) as specified around 
Eq.~(\ref{H0R}) and below (\ref{HtildeR}).
(b): Square lattice models (iii) and (iv) as specified around 
Eq.~(\ref{H02D}) and below (\ref{HtildeR}).
The grey and white regions indicate our choice of the
sublattices $\Lambda_1$ and $\Lambda_2$ in (\ref{HtildeR}).
All the remaining model parameter values in (\ref{psi0}) and
(\ref{HtildeR})
have been chosen in (a) as detailed in Fig.~\ref{fig2},
and in (b) as detailed in Fig.~\ref{fig3}.
}
\label{fig1}
\end{figure}

As our initial condition $\rho(0)$ (see above Eq.~(\ref{8a}))
we choose a pure state of the form 
$\rho(0)=|\psi\rangle\langle\psi|$ with
\begin{eqnarray}
|\psi \rangle \propto e^{-\frac{\beta}{2} \tilde H}| \phi\rangle
\ ,
\label{psi0}
\end{eqnarray}
where $|\phi\rangle$ is a normalized random vector, 
which may be viewed as point on the unit 
sphere in $\CC^{(2s+1)^{\kappa}}$,
randomly sampled according to a uniform 
distribution.
It is well-known that such an initial condition 
exhibits a so-called dynamical typicality property,
meaning that it imitates very accurately the behavior 
of the canonical ensemble from \eqref{23}, see, e.g. 
Ref. \cite{rei} and further references therein.
More precisely speaking, for the vast majority of
all those randomly sample initial states
$\rho(0)=|\psi\rangle\langle\psi|$,
the time-dependent expectation values in 
(\ref{8a}) become, for sufficiently large 
system sizes $\kappa$, practically indistinguishable 
from those which one would obtain by choosing 
$\rho(0)$ according to (\ref{23}).
A more precise analytical quantification of 
the remaining deviations
is in general
quite difficult, but we numerically verified 
that our results for different random initial states
were indeed nearly indistinguishable
on the scale of the subsequent plots.
Apart from this connection to the canonical ensemble in (\ref{23}),
our initial state (\ref{psi0}) represents, of course, already
in itself a perfectly legitimate, generally far from 
equilibrium initial condition.

\begin{figure}
\includegraphics[scale=1.3]{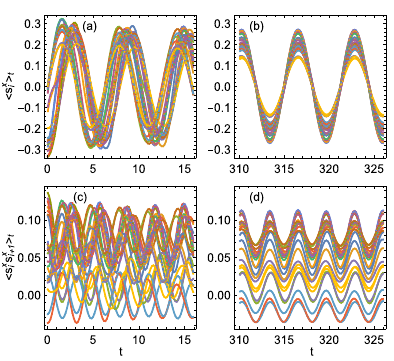}
\caption{
(a) and (b): Expectation values \eqref{8a} of the local 
observables $A=s_i^x$ for early times (a) as well as for late times (b)
by numerically solving the spin ring model from \eqref{H0R} with 
$\kappa=24$ spins, periodic boundary conditions,
random couplings $J_i \in \lbrack-3,1\rbrack$,
and magnetic field $h = 1$, see also Eqs.~\eqref{1}-\eqref{3}. 
The different colors correspond to the 24 different 
observables $A=s_i^x$.
The 
initial condition $\rho(0)$
is given by a 
canonical ensemble of the form \eqref{23}, \eqref{24} with 
$\beta=1$, choosing the Hamiltonian $\tilde H$ according to 
\eqref{HtildeR} with $\tilde H_0=H_0$,
$h_x=h_y=1$,
and sublattices $\Lambda_{1,2}$ as indicated in Fig.~\ref{fig1}(a),
see also main text for more details.
In the actual numerics, the behavior of the corresponding time evolved 
$\rho(t)$ was imitated by numerically evolving a random initial state 
as explained around Eq.~\eqref{psi0}.
(c) and (d): Same, but for the observables $A=s_i^xs_{i+1}^x$ with $i=1,...,23$.
}
\label{fig2}
\end{figure}

\begin{figure}
\includegraphics[scale=1.3]{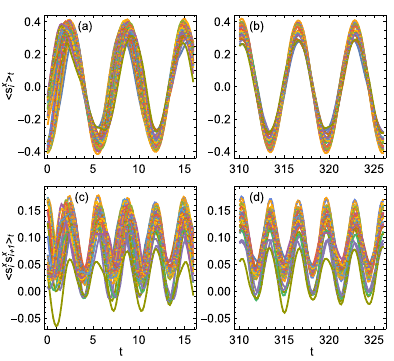}
\caption{
Same as in Fig.~\ref{fig2}, but now for a $5\times 5$ square lattice model 
of the form \eqref{H02D} with $\kappa=25$ spins, open boundary conditions,
and couplings $J=-2$.
In particular, the initial condition is again of the form
\eqref{23}, \eqref{24}, \eqref{HtildeR}
with $\beta=1$, $\tilde H_0=H_0$,
and 
sublattices $\Lambda_{1,2}$ as indicated in Fig.~\ref{fig1}(b).
}
\label{fig3}
\end{figure}

Once the initial state has been chosen,
we numerically evolved it in time 
by means of Suzuki-Trotter product expansion techniques,
as detailed, for instance, in Ref. \cite{michi}.

While this temporal evolution is governed by the
above specified, so-called post-quench Hamiltonian $H$
(see also Sec. \ref{s33}), 
the so-called pre-quench Hamiltonian $\tilde H$, 
governing the initial condition via
(\ref{23}) and (\ref{psi0}), is 
chosen as
\begin{eqnarray}
\tilde H & := & \tilde H_0  
+ 
h_x
\sum_{i\in \Lambda_1} s_i^x 
+ 
h_y
\sum_{i\in \Lambda_2} s_i^y
\ ,
\label{HtildeR}
\end{eqnarray}
where $\tilde H_0$ is 
of the same general 
structure as in (\ref{H0R}) in our one-dimensional examples 
(i) and (ii), and as in (\ref{H02D}) in our two-dimensional 
examples (iii) and (iv).
More precisely speaking, $\tilde H_0$ was chosen identical 
to $H_0$ from (\ref{H0R}) and (\ref{H02D}) in (i) and (iii), 
respectively,
while the same $\tilde H_0$'s as in (i) and (iii) were 
then also 
employed in (ii) and (iv), respectively.
Furthermore, $\Lambda_1$ and $\Lambda_2:=\Lambda\setminus\Lambda_1$ 
in (\ref{HtildeR}) 
denote two complementary subsets of the respective
total lattices $\Lambda$ (see above Eq.~(\ref{1})).
Their specific choice for the examples
(i) and (ii) is visualized by the grey and white regions
in \figref{fig1}(a), and for the examples (iii) and (iv) 
in \figref{fig1}(b).
According to (\ref{HtildeR}), the spins
in those two sublattices (grey and white)
are thus polarized by the external magnetic 
fields $h_x$ and $h_y$ along orthogonal directions, 
resulting via (\ref{psi0}) in initial conditions
for the individual spins as cartooned by the red 
arrows in \figref{fig1}.

Such inhomogeneous initial states with 
two extended domains of macroscopic magnetization 
appeared to us as particularly interesting and 
non-trivial examples.
For instance, they clearly are {\em not} translationally 
invariant (see also Sec.~\ref{s4} below).
Moreover, they are far from thermal equilibrium with respect
to the post-quench Hamiltonian $H$.

As a first example, \figref{fig2} displays 
the dynamics of various local observables for a spin 
ring model of type (i) with parameters given in the caption. 
With these parameters, the energy is not close 
to the edges of the spectrum. 
At early times, panels (a) and (c), 
the different observables behave
rather irregularly, starting from their various initial values, 
whereas at later times, panels (b) and (d), all observables 
exhibit quite regular oscillations with angular frequency
$h$ in (b) and $2h$ in (d),
thus confirming and illustrating our main
analytical prediction from the previous subsections.
The phases between these long-time oscillations 
seem to be astonishingly small, but the amplitudes 
differ quite notably (and in (d) also the 
time-averaged values).
We will briefly return to this observation
at the end of Sec.~\xref{s4}.

For the two-dimensional square lattice model of type (iii),
qualitatively quite similar results are observed in
\figref{fig3}. 
The main difference is that some of the long-time oscillations,
especially in (d), still exhibit notable deviations from a 
strictly periodic behavior,
which can be naturally understood as finite-size 
corrections to our analytical predictions.
A more detailed discussion of these finite-size
effects is provided in Appendix \ref{app3}.

\begin{figure}
\includegraphics[scale=1.3]{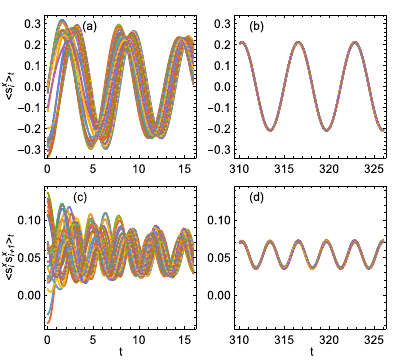}
\caption{
Same as in Fig.~\ref{fig2}, but now for non-random 
couplings $J_i=-1$ in the spin ring model (\ref{H0R}).
In particular, exactly the same the initial condition 
as in Fig.~\ref{fig2} was utilized.
}
\label{fig4}
\end{figure}

Turning to the two remaining, translationally invariant models (ii) and (iv),
we encounter almost perfect synchronization of the individual local 
observables after initial transients have died out.
Figure~\xref{fig4} displays this behavior for
the spin ring model (ii), where local spin operators are related 
to each other by a translation along the ring, 
a symmetry operation under which the Hamiltonian is invariant,
see also Sec. \ref{s4} below.
For late times, panels (b) and (d), the various oscillations 
superimpose perfectly, although the initial state is exactly the same
as 
the one in \figref{fig2}, i.e.\ adapted to the Hamiltonian
with disorder. We have verified that this synchronization 
behavior is practically independent of the initial conditions.

\begin{figure}
\includegraphics[scale=1.3]{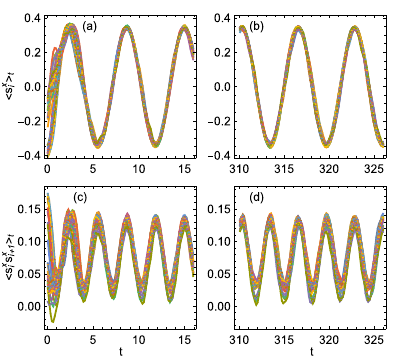}
\caption{
Same as in Fig.~\ref{fig3}, but now for periodic 
boundary conditions in the $5\times 5$
square lattice model (\ref{H02D}).
In particular, exactly the same initial condition 
as in Fig.~\ref{fig3} was utilized.
}
\label{fig5}
\end{figure}

A qualitative similar behavior is also observed for
our $5\times 5$ square lattice model (iv)
in \figref{fig5}.
As will be explained in more detail in Sec.~\ref{s4},
the observed synchronization at large times
has its origin in the model's translational 
invariance.
Similarly as in Fig.~\ref{fig3}, the remnant deviations
from prefect synchronization, especially in Fig.~\ref{fig5}(d),
can be explained in terms of finite-size effects.
Apparently, the fact that our square lattice models (iii) and (iv)
only exhibit a relatively short period of $5$ along each spatial 
direction is responsible for the stronger finite-size
corrections in comparison to the spin ring models (i) and (ii).
Similarly as in Appendix \ref{app3},
we also confirmed this expectation by directly comparing 
the numerical results for different system sizes with each 
other (not shown).

Further numerical examples for a variety of other model Hamiltonians 
and, more importantly, other initial conditions can also be 
found in Ref.~\cite{vor21}.

\section{Synchronization}
\label{s4}

A particularly remarkable feature of the numerical results in 
Figs.~\ref{fig4} and \ref{fig5} is the close agreement of all 
the differently colored graphs for sufficiently late times (right panels),
while Figs.~\ref{fig2} and \ref{fig3} do {\em not} exhibit 
such a behavior. In the following, our main objective is a 
better understanding of this numerical observation.

For the sake of simplicity, we mainly focus
on one-dimensional spin models (\ref{2}). 
Moreover, we require that the model is
translationally invariant in the sense that
site $i=\dof+1$ is identified with $i=1$ (periodic boundary conditions)
and the couplings $J_{ij}$ only depend on the  difference $i-j$ modulo $\dof$.
Finally, we restrict ourselves to
the generic case that all energies $E_n^0$ 
are pairwise different,
hence the quantities $\bar f_\nu$ are given by (\ref{16}),
see also the remarks below Eqs.~(\ref{6}) and (\ref{16}).

For the rest, short- as well as long-range interactions are still admitted.
Moreover, various generalizations, e.g., to higher-dimensional 
hypercubic lattices (with periodic boundary conditions) are straightforward,
see also Sec.~\ref{s21}, but will not be explicitly worked out.

Denoting by $\TT$ the so-called translation operator,
it is shown in Appendix \ref{app2} that
\begin{eqnarray}
\langle n, l |\TT^\dagger\! B \TT |n,l'\rangle=\langle n, l | B | n,l' \rangle
\label{28}
\end{eqnarray}
for arbitrary Hermitian operators $B$ and 
indices
$n,l,l'$.
Physically, $\TT^\dagger\! B \TT$ represents the same observable
as $B$, except that ``everything is shifted'' by one unit along the 
periodic spin chain.
For instance, for a single-site spin operator $s_i^a$ 
(with $a\in\{x,y,z\}$)
one finds that $\TT^\dagger\! s_i^a \TT=s_{i+1}^a$, 
and analogously for arbitrary sums and products of such operators.
In particular, the total spin components from (\ref{1}) 
and the Hamiltonian from (\ref{2}) are found to be
translationally invariant  in the sense that they commute with $\TT$.

Taking into account (\ref{11}) and (\ref{28}),
one can conclude that the quantities $\bar f_\nu$ in (\ref{16}) 
and thus the function $\Att$ in (\ref{14}) and (\ref{25})
remain unchanged if we replace the observable $A$ by 
its shifted counterpart $\TT^\dagger\! A \TT$.
For instance, the expectation values of the single-site 
spin operators $s_i^a$ are thus
predicted to synchronize (look the same for all $i$)
in the long run, 
and likewise for arbitrary sums and products of 
such operators.
These findings are illustrated by 
Figs. \ref{fig4} and \ref{fig5},
see also Sec.~\xref{s35}.
The small remnant deviations from strict synchronizations
in these numerical examples can be naturally understood as
finite-size effects,
see also Appendix \ref{app3}.

It readily follows that so-called local operators $A_i$ with 
the property $\TT^\dagger\! A_i \TT=A_{i+1}$
will synchronize in the above sense not only with each other
but also with their ``intensive'' counterpart
$A:=\sum_{i\in \G } A_i/\dof$, as exemplified by Eq.~(\ref{29}) below.
We also remark that all these conclusions apply
to arbitrary initial states $\rho(0)$ 
(as long as they satisfy (\ref{20})).
In particular, $\rho(0)$ is {\em not} 
required to be translationally invariant.

An analogous line of reasoning implies that
the initial condition $\rho(0)$ and its shifted counterpart
$\TT^\dagger\! \rho(0) \TT$ exhibit in the long run
(nearly) identical expectation values for arbitrary 
observables $A$ and any initial state $\rho(0)$ 
which satisfies (\ref{20}).

Altogether, the generically occurring, permanent long-time oscillations 
from Sec.~\ref{s34} are thus found to synchronize in the sense of being 
invariant under arbitrary translations of the considered observable, 
provided the system Hamiltonian 
(but not necessarily the initial condition) is translationally invariant.

Closely related numerical findings have been recently reported in Ref.~\cite{vor21}.
Our present work 
amounts to a rigorous analytical validation and 
generalization of this numerical discovery of synchronization
in closed
systems of the form (\ref{1})-(\ref{3}).
Similarities and differences with respect to related 
synchronization phenomena in 
{\em open} 
systems have also
been addressed already in Ref.~\cite{vor21} (see also \cite{buc22}),
and are therefore not repeated here.
The salient point is that while the observable phenomena are
similar, the basic physical mechanisms 
as well as the analytical methods are entirely different 
for closed and open systems.

Various slightly different 
notions of synchronization 
are reviewed, for instance, in Ref.~\cite{buc22}.
Our present notion appears to us particularly simple and natural.

Intuitively, and also on the basis of our above calculations, it
seems reasonable to suspect that translational invariance, 
or equivalence of all spin sites in general, 
is not only sufficient but that it generically is
even necessary for the occurrence of synchronization 
in our present sense.
This expectation is further corroborated by the numerical examples 
in Figs.~\xref{fig2} and \xref{fig3}.

\section{Simple Analytical examples}
\label{s5}

Of foremost interest are cases where
$\bar f_\nu$ in (\ref{15}) is non-zero at least for one $\nu\not=0$, 
giving rise to non-equilibration in the form of permanent 
oscillations in 
(\ref{25}).
In general, the explicit evaluation of
$\bar f_\nu$ in (\ref{15}) is a quite demanding task.
In the following, we focus on some particularly
simple examples.

\subsection{Single spins}
\label{s51}

To begin with, we illustrate the general idea by means 
of the observables
\begin{eqnarray}
M_a:=\frac{1}{\dof}\, S^a=\frac{1}{\dof}\sum_{i\in \G }s_i^a
\ ,
\label{29}
\end{eqnarray}
see also Eq.~(\ref{1}),
i.e., the $M_a$ are essentially the magnetizations
along the spatial direction $a\in\{x,y,z\}$.
Employing the usual 
raising and lowering operators
\begin{eqnarray}
S^\pm & := & S^x\pm i S^y
\label{30}
\end{eqnarray}
one readily recovers 
the 
relations
(see also around Eqs.~(\ref{4})-(\ref{6}))
\begin{eqnarray}
S^\pm |n,l\rangle & = & c^\pm_{n,l} |n,l\pm 1\rangle
\label{31}
\\
c^\pm_{n,l} & := & \sqrt{L_n(L_n+1) - l (l\pm 1)}
\ .
\label{32}
\end{eqnarray}

Observing that Eqs.~(\ref{12}) and (\ref{13}) are linear in $A$, 
the same equations must also apply to the non-Hermitian operator 
$A:=S^+$ from (\ref{30}).
Exploiting (\ref{11}) and (\ref{31}), it follows that
\begin{eqnarray}
A_{nm}^{k+\nu,k} = \delta_{n,m}\, \delta_{\nu,1} \, c_{n,k}^+
\ ,
\label{33}
\end{eqnarray}
where $\delta_{n,m}$ and $\delta_{\nu,1}$ are Kronecker deltas.
Hence, we can conclude with Eqs.~(\ref{13}) and (\ref{15}) that
\begin{eqnarray}
f_{\nu}(t) = \bar f_\nu = \delta_{\nu,1}\, f_1(0)
\label{34}
\end{eqnarray}
and with 
Eq.~(\ref{14}) and (\ref{12}) that 
$\Att:= \At= \langle A\rangle_{\! 0}e^{iht}$.
By means of a similar line of reasoning for $A:=S^-$ one thus
arrives at
\begin{eqnarray}
{\cal S}_t^\pm
:=\langle S^\pm\rangle_{\! t} 
= \langle S^\pm\rangle_{\! 0} e^{\pm iht}
\ .
\label{35}
\end{eqnarray}

Since $S^x=(S^+ + S^-)/2$ and $S^y=(S^+ - S^-)/2i$ according to
(\ref{30}), we finally obtain for the magnetizations $M_{x,y}$ from (\ref{29})
the result
\begin{eqnarray}
\langle M_x\rangle_{\! t} 
& = & 
a_1 \cos (ht) - b_1 \sin(ht)
\ ,
\label{36}
\\
\langle M_y\rangle_{\! t} 
& = & 
b_1 \cos (ht) + a_1 \sin(ht)
\label{37}
\\
a_1
& := & 
\langle M_x\rangle_{\! 0}
\ , 
\label{38}
\\
b_1 
& := & 
\langle M_y\rangle_{\! 0}
\ . 
\label{39}
\end{eqnarray}
i.e., these particular observables 
exhibit perfect harmonic oscillations 
for all times $t$ and
for any initial state
$\rho(0)$ with a non-vanishing expectation 
value of $M_x$ or of $M_y$.
In the same way one finds that
\begin{eqnarray}
\langle M_z\rangle_{\! t} 
& = & 
\langle M_z\rangle_{\! 0} 
\ ,
\label{40}
\end{eqnarray}
i.e., this particular observable is, as expected, always
a conserved quantity.

Similarly as in the first equality in (\ref{35}), one readily sees 
that for the three specific observables $A:=M_a$ from above,
the auxiliary functions $\Att$ 
happen to be exactly identical to the
true expectation values $\At$ 
for all $t$.
In other words, none of further preconditions on the
energies $E_n^0$, the system size $\dof$, 
and the initial condition $\rho(0)$
from Sec.~\ref{s32} are actually
needed in these specific examples.

In a next step, let us focus on systems which
satisfy the preconditions for our main result
in Sec.~\ref{s32} as well as the preconditions 
for synchronization as detailed at the beginning 
of Sec.~\ref{s4}.
According to the discussion at the end of
Sec.~\ref{s4}, we thus can conclude that
the single-spin expectation values
$\langle s_i^a\rangle_{\! t}$ 
behave for most sufficiently large times $t$ very similarly
to each other and thus to 
$\langle M_a\rangle_{\! t}$ (see Eq. (\ref{29})),
symbolically indicated as
\begin{eqnarray}
\langle s_i^a\rangle_{\! t}
\rightsquigarrow
\langle M_a\rangle_{\! t}
\ ,
\label{41}
\end{eqnarray}
where $a\in\{x,y,z\}$.
In particular, for any given observable $A:=s_i^x$,
the corresponding auxiliary function $\Att$
takes the $i$-independent explicit form (\ref{36}),
and similarly for $s_i^y$ and $s_y^z$.
On the other hand, for short times $t$ the expectation values
$\langle s_i^a\rangle_{\! t}$ are in general
no longer close to $\langle M_a\rangle_{\! t}$.
Rather, and as can be seen in Figs. \ref{fig4} and \ref{fig5},
any given $\langle s_i^a\rangle_{\! t}$ 
generically exhibits a non-trivial initial
relaxation process of its own, whose details depend 
in a complicated manner on the initial state $\rho(0)$ 
and on the Hamiltonian $H$.
Moreover, even for large times $t$ there will 
generically remain
fluctuations of $\langle s_i^a\rangle_{\! t}$ about
$\langle M_a\rangle_{\! t}$, which are negligibly small
for most $t$ but may become large for some
very rare $t$'s (quantum recurrences or revivals \cite{rid23}).

\subsection{Higher harmonics}
\label{s52}

Our next examples are observables of the form $A:=M_a^2$.
By means of similar calculations as before 
one finds  that
\begin{eqnarray}
\langle M_x^2\rangle_{\! t} 
& = & 
a_2\, \cos (2ht) - b_2 \sin(2ht) + c_2
\ ,
\label{42}
\\
\langle M_y^2\rangle_{\! t} 
& = & 
-a_2\, \cos (2ht) + b_2 \sin(2ht) + c_2
\ ,
\label{43}
\\
\langle M_z^2\rangle_{\! t} 
& = & 
\langle M_z^2\rangle_{\! 0} 
\ ,
\label{44}
\end{eqnarray}
where we introduced the abbreviations
\begin{eqnarray}
a_2
& := & 
\langle M_x^2-M_y^2\rangle_{\! 0}/2
\ , 
\label{45}
\\
b_2 
& := & 
\langle M_xM_y+M_yM_x\rangle_{\! 0}/2
\ . 
\label{46}
\\
c_2
& := & 
\langle M_x^2+M_y^2\rangle_{\! 0}/2
\ , 
\label{47}
\end{eqnarray}
As expected,  $M_x^2+M_y^2$ and $M_z^2$ are 
thus conserved quantities.
Moreover, the observables $M_{x,y}^2$ exhibit perfect harmonic 
oscillations for all $t$ and for all initial conditions $\rho(0)$
with a non-vanishing expectation value in (\ref{44}) or in (\ref{46}).
Last but not least, the oscillation frequency is now twice as large 
as in  (\ref{36}) and (\ref{37}) (higher harmonics).

Combining (\ref{36}) and (\ref{42}) one can conclude that
\begin{eqnarray}
\langle M_x^2\rangle_{\! t} - \langle M_x\rangle^2_{\! t}
\!\!
& = & 
\!\!
a_2' \cos (2ht) 
- 
b_2' \sin(2ht) 
+ c_2'
\, ,
\label{48}
\\
a_2'
& := &
(\sigma^2_{xx}-\sigma^2_{yy})/2
\ ,
\label{49}
\\
b_2'
& := &
(\sigma^2_{xy}+\sigma^2_{yx})/2
\ ,
\label{50}
\\
c_2'
& := &
(\sigma^2_{xx}+\sigma^2_{yy})/2
\ ,
\label{51}
\end{eqnarray}
where
\begin{eqnarray}
\sigma^2_{ab}
& := & 
\langle M_a M_b\rangle_{\! 0} 
- 
\langle M_a \rangle_{\! 0}
\langle M_b \rangle_{\! 0} 
\label{52}
\end{eqnarray}
for arbitrary $a,b\in\{x,y,z\}$.
Analogous results as for $M_x$ in (\ref{48})
apply to $M_y$ and $M_z$.

Incidentally, for observables of the form
$s_i^xs_j^x$ one still can deduce from
(\ref{5}) and (\ref{15}) 
that the long-time asymptotics must be of 
the general structure
\begin{eqnarray}
\langle s_i^xs_j^x\rangle_{\! t}
\rightsquigarrow
a_{ij} \cos (2ht) 
+ b_{ij} \sin(2ht) 
+ c_{ij}
\ .
\label{53}
\end{eqnarray}
Most importantly, the oscillation frequency is 
again twice as large as in (\ref{36}), (\ref{41}),
in accordance with the numerical examples in Figs. 
\ref{fig2}-\ref{fig5}.
Similarly as in Eqs.~(\ref{36})-(\ref{41}), 
the coefficients $a_{ij}, b_{ij}, c_{ij}$ in (\ref{53})
are once more independent of $h$,
but now their quantitative dependence on the 
initial state $\rho(0)$ and on the Hamiltonian $H_0$ is 
very difficult to specify in more detail.
Analogous statements apply to observables of the form 
$s_i^a s_j^b$ with $a,b\in\{x,y\}$ and to products of more 
than two such factors.

\subsection{Thermodynamic limit}
\label{s53}

Next we turn to the issue of how the above findings 
depend on the system size $\dof$, and, in particular,
how they behave for asymptotically large $\dof$, 
i.e.\ in the thermodynamic limit.
As usual in this context, we focus on systems whose 
size can be ``upscaled'' in a physically natural way. 
Particularly simple examples are translationally
invariant Hamiltonians (see Sec.~\ref{s4})
with short-range interactions, i.e., the couplings
$J_{ij}$ in (\ref{3}) decay sufficiently fast 
(and independent of $\dof$) with increasing distance 
between the two sites $i$ and $j$.
Similarly, the initial states $\rho(0)$ must be chosen so
that they amount to ``physically similar situations''
for different system sizes $\dof$.
For example, the system energy $\tr\{\rho(0)H\}$
is often expected to grow linearly with the 
system size $\dof$,
i.e., the energy density (energy per degree of freedom) 
is kept constant.
Simple examples are canonical ensembles
of the form (\ref{23}), (\ref{24}) with fixed
parameters $\beta$ and $\vec h_i$
(independent of $\dof$ and $i$).

Rather than trying to formally define this class of
``extensive'' Hamiltonians $H$ and initial states 
$\rho(0)$ more precisely, we assume as
a ``minimal requirement'' that the concomitant
expectation values of ``intensive observables'',
such as the magnetization $M_a$ in
(\ref{29}), can be considered as asymptotically 
independent of the system size $\dof$,
and that their statistical fluctuations and/or quantum 
uncertainties, as exemplified by (\ref{52}),
decay to zero with increasing system size $\dof$
(usually as $1/\dof$).
Moreover, we assume that correlations between local
observables in the initial state $\rho(0)$,
such as
\begin{eqnarray}
c_{ij}^{ab}:=\langle s_i^a s_j^b \rangle_{\! 0}
- \langle s_i^a\rangle_{\! 0}\langle s_j^b \rangle_{\! 0}
\ ,
\label{54}
\end{eqnarray}
decay to zero with the distance between the sites $i$ and $j$
sufficiently fast and asymptotically independently
of the system size $\dof$.
Essentially, this assumption
is tantamount to the so-called
cluster decomposition property \cite{wic63,wei97,ess16,mur19,glu19}.
Though this property has until now only be rigorously 
established for a quite restricted set of examples
\cite{ara69,par82,par95,kli14,fro15}, it
is commonly expected to be obeyed 
by any ``physically realistic'' $\rho(0)$
-- at least outside the 
realm where phase transitions may occur.

In particular, for systems that
possibly may exhibit large thermal fluctuations
as a precursor of spontaneous symmetry breaking
in the thermodynamic limit, the 
energy density must be chosen 
outside the range where such effects 
occur.
The opposite situation will be further explored in Sec.~\ref{s6}.

Given that the initial magnetizations $\langle M_a \rangle_{\! 0}$
are asymptotically independent of the system size $\dof$,
the same follows for any later time $t$ 
according to (\ref{36})-(\ref{40}), and thus
for the late-time behavior of any single spin 
according to (\ref{41}).

In the same vein, the initial expectation values in (\ref{45})-(\ref{47})
are expected to be asymptotically independent of the system size
$\dof$ for physically realistic initial states $\rho(0)$, hence the same
applies to the time-dependent expectation values in (\ref{42})-(\ref{44}).
On the other hand, the initial variances $\sigma^2_{aa}$ 
(see (\ref{52}))
generically decay to zero for large $\dof$.
The same follows for the correlations $\sigma^2_{ab}$ 
in (\ref{52})
upon observing that
$[\sigma^2_{ab}]^2\leq \sigma^2_{aa}\sigma^2_{bb}$ 
(Cauchy-Schwarz inequality), and hence for the
variance of $M_x$ in (\ref{48}), and similarly for $M_y$ and $M_z$.
Essentially, this reflects the common fact that quantum and statistical
fluctuations become negligible for macroscopic observables.
The main conclusion is that $\langle M_x^2\rangle_{\! t}$
can often be very well approximated by $\langle M_x\rangle^2_{\! t}$.

Finally it is reasonable to expect that a large-$\dof$ asymptotics
qualitatively similar to (\ref{41}) will also apply to local 
observables of the form $s_i^a s_j^b$. 
However, more rigorous and/or quantitative 
statements along these lines are difficult to obtain,
see also the discussion below Eq.~(\ref{53}).

On the other hand, quantum and statistical
fluctuations of microscopic (local) observables 
are well-known to generically remain non-negligible.
Accordingly, 
correlations at the initial time $t=0$,
as exemplified by (\ref{54}),
with not too large distances between the sites $i$ and $j$,
are not expected to approach zero for large $\dof$, and likewise 
for the analogous correlations at any later time point $t$.
Numerical examples in support of this expectation are provided 
by Figs. \ref{fig2}-\ref{fig5}.

\subsection{Final remarks}
\label{s54}

Our first remark is that in case of the macroscopic
observables (\ref{29}), the exact time-dependencies 
(\ref{36})-(\ref{40})
can also be obtained ``directly'', i.e., without 
exploiting our main results from Sec.~\ref{s3},
and likewise for (\ref{42})-(\ref{47}).
Namely, by exploiting the specific symmetries of 
the Hamiltonian $H$ in (\ref{2}), the Heisenberg 
equations of motion
which govern the expectation values of 
those observables can be readily solved, as 
detailed, e.g., in Ref \cite{vor21}.
From this viewpoint, the absence of equilibration 
in such models may thus be considered
as a relatively obvious consequence of their special
symmetry properties.

For most other observables, the generic occurrence of 
permanent long-time oscillations is a far from obvious
key finding of our present work.
The fact that this finding is indeed non-trivial is
already quite evident by recalling that usually 
an (approximately) time-periodic behavior only appears 
after sufficiently long times (see Figs. \ref{fig2}-\ref{fig5}), 
and even then the actual expectation values still 
exhibit certain deviations from strict periodicity
(for systems of finite size).
Moreover, the oscillations are asynchronous 
unless the system happens to be translationally 
invariant (Sec.~\ref{s4}).

Our second remark is that ``single spin observables''
$s_i^a$ with $a\in\{x,y\}$ and their intensive 
counterparts $M_a$ from (\ref{29})
were found in Sec.~\ref{s51}
to exhibit 
harmonic long-time oscillations 
with angular frequency $h$.
In the same vein, ``two-spin observables''
$s_{i_1}^{a_1}s_{i_2}^{a_2}$ with $a_{1,2}\in\{x,y\}$
were found to harmonically oscillate with angular 
frequency $2h$ in Sec.~\ref{s52},
while $\langle M_a^2\rangle_{\! t}$ turned out to 
be often close to $\langle M_a\rangle^2_{\! t}$ 
in Sec.~\ref{s53}.
Analogously, it is quite evident that harmonic 
oscillations with angular frequency $\nu h$ will 
arise for $\nu$-spin observables
$s_{i_1}^{a_1}\cdots s_{i_\nu}^{a_\nu}$,
while
$\langle M_a^\nu\rangle_{\! t}$ will be close to
$\langle M_a\rangle^\nu_{\! t}$ in many cases.
The latter example implies that the long-time 
oscillations are in general {\em not} of a 
purely harmonic character.

\section{Equilibrium correlations and time crystals}
\label{s6}

Throughout this section we restrict ourselves to
system states of the specific form
\begin{eqnarray}
\rho = \sum_{nl} p_{nl}\, |n,l\rangle\langle n,l |
\label{55}
\end{eqnarray}
with $p_{nl}\geq 0$ and $\sum_{nl} p_{nl}=1$.
It follows from (\ref{7}) that $[H,\rho]=0$, i.e. the 
state $\rho$ remains unchanged in the course of time
(steady or equilibrium state).
Particularly important examples are 
thermal equilibrium 
ensembles
of the canonical form
\begin{eqnarray}
\rho = e^{-\beta H}\!/\tr\{e^{-\beta H}\}
\ .
\label{56}
\end{eqnarray}
Other examples are microcanonical ensembles, or, more generally,
largely arbitrary diagonal ensembles of low purity, 
see also Eqs.~(\ref{22}), (\ref{26}), and below Eq.~(\ref{61}).

In other words, we are dealing here with the 
exceptional cases announced below Eq. (\ref{25}), 
for which any permanent oscillations are 
strictly ruled out.
Our main objective in this section is to show that 
the basic SU(2) symmetry (see above 
Eq. (\ref{4})), which is at the origin
of the permanent oscillations in the 
generic case, still gives rise to some
different kind of interesting properties
in our present exceptional cases,
including systems at thermal equilibrium 
as particularly prominent examples.
In order to achieve this goal, the salient 
point will be to consider so-called temporal 
correlations (see Eq. (\ref{57}) below) 
instead of the so-far employed expectation 
values (see Eq. (\ref{8a})).
Incidentally, these explorations will at the same 
time very naturally open up a connection to the
topic of time crystals, which recently attracted
a considerable amount of attention.

As announced, the quantities of foremost interest 
throughout the present section will be
temporal correlations,
also called, among others, dynamic or 
two-point correlation functions,
and being formally defined as
\begin{eqnarray}
C_{\! AB}(t) :=\tr\{\rho A B(t)\}
\label{57}
\end{eqnarray}
for any given pair of observables $A$ and $B$,
where $B(t):=e^{iHt}Be^{-iHt}$ (Heisenberg picture, $\hbar=1$).

Similarly as in (\ref{9})-(\ref{13}) one finds that
\begin{eqnarray}
C_{\! AB}(t) 
& \!\! = \!\! &  
\sum_\nu g_{\nu}(t) \, e^{i\nu h t}
\ ,
\label{58}
\\
g_{\nu}(t) 
& \!\! := \!\! & 
\sum_{mn} e^{i(E_n^0-E_m^0)t}\sum_k  p_{mk}\, A_{mn}^{k,k+\nu} B_{nm}^{k+\nu,k}
\, , \ \ 
\label{59}
\end{eqnarray}
and similarly as in (\ref{14}), (\ref{15}), (\ref{21}) that
\begin{eqnarray}
C_{\! AB}(t) 
&  \rightsquigarrow &
\sum_\nu \bar g_{\nu} \, e^{i\nu h t}
\ ,
\label{60}
\\
\bar g_{\nu}
& := &
{\sum_{mnk}}' p_{mk}\, A_{mn}^{k,k+\nu} B_{nm}^{k+\nu,k}
\label{61}
\end{eqnarray}
under the very same preconditions as those discussed 
in Secs.~\ref{s32} and \ref{s33}.
The detailed derivation is quite similar to Appendix \ref{app1}
-- see also Supplemental Material of Ref.~\cite{alh20} --
and therefore omitted here.

As a consequence, the generic 
appearance 
of permanent oscillations is predicted similarly as
in Sec.~\ref{s34}, and of synchronization effects 
similarly as in Sec.~\ref{s4} in case of translationally 
invariant systems.
In particular, correlations of local observables
$A_i$ and $B_i$ with the property $\TT^\dagger\! A_i \TT=A_{i+1}$
and $\TT^\dagger\! B_i \TT=B_{i+1}$
are predicted to synchronize with each other,
and also with the correlations of their intensive
counterparts
$A:=\sum_{i\in \G } A_i/\dof$ and $B:=\sum_{i\in \G } B_i/\dof$,
respectively.

Note that the correlation in (\ref{57}) is, in general,
a complex valued function
of $t$, and as such not an immediately 
observable quantity. However, analogous predictions
readily carry over to its
real (symmetrized) part
\begin{eqnarray}
C^s_{\! AB}(t) := 
[\tr\{\rho A B(t)\}+\tr\{\rho B(t) A\}]/2
\ ,
\label{62}
\end{eqnarray}
and analogously for its imaginary part.

Focusing on the specific observables $A=B=M_x$ from
(\ref{29}), one finally finds, similarly as in Sec.~\ref{s5}, 
for arbitrary $t$ and without any further approximation 
that
\begin{eqnarray}
C^s_{M_x\!M_x}(t) & = &
\tilde a_2
\, \cos(ht)
\ ,
\label{63}
\\
\tilde a_2 & := & \tr\{\rho M_x^2 \}
\ ,
\label{64}
\end{eqnarray}
and likewise for $A=B=M_y$.
In case of a translationally invariant system, 
we furthermore can conclude under similar conditions as 
above (\ref{41}) that
\begin{eqnarray}
C^s_{s_i^x\! s_i^x}(t)
\rightsquigarrow
C^s_{M_x\!M_x}(t) 
\ .
\label{65}
\end{eqnarray}

These findings imply interesting conclusions
with respect to the topic of time crystals.
At the focus of the latter issue are, generally speaking, 
various conceivable forms and disguises 
of a spontaneously broken time-translation symmetry,
see, e.g., Refs.~\cite{han22,ven19} for recent reviews.
Here, we specifically address the possible occurrence 
of such fascinating phenomena 
in {\em isolated} many-body quantum systems 
{\em at thermal equilibrium},
meaning that
no periodic driving and no external bath(s) or other
sources of dissipation are involved, nor do we focus
on the zero temperature limit or ground state properties,
nor is the thermodynamic 
limit taken before the long-time limit \cite{han22,ven19}.
Under these circumstances, a particularly well-established 
definition of a time crystal explicitly refers to the behavior of
temporal correlations at thermal equilibrium, 
requiring that they must exhibit permanent oscillations 
in time as well as long-range order in space \cite{wat15}.
In our present context, this is largely equivalent \cite{wat15,wat20,hua19}
to the requirement that there must exist intensive observables $A,B$,
as exemplified by (\ref{29}) and more generally defined below Eq.~(\ref{61}),
whose correlation function in (\ref{57})
exhibits permanent oscillations that do not
tend to zero for asymptotically large system
size $\dof$, see below (\ref{54}).

Combining this definition and Eq.~(\ref{63}), a time crystal will thus be 
realized by focusing on the example $A=B=M_x$ and showing
that $\tilde a_2$ in (\ref{64}) approaches a positive limiting value
for asymptotically large $\dof$ in the canonical ensemble from
(\ref{56}).
Observing (\ref{64}) and that $\tr\{\rho M_x\}=0$
such a behavior of $\tilde a_2$ is tantamount
to the appearance of macroscopic thermal fluctuations
of $M_x$ and is thus expected to 
arise
if the Heisenberg model in (\ref{1})-(\ref{3}) exhibits 
in the thermodynamic limit a spontaneous symmetry 
breaking (phase transition) with respect to $M_x$.
In this context it may be worth to recall that, 
as always, we tacitly focus on cases with a
non-vanishing external field $h$ in (\ref{2}).

Remarkably, we thus established a direct
connection between a spontaneously broken time-translation
invariance in the context of time crystals, and a
spontaneously broken spatial symmetry in the context
of phase transitions at thermal equilibrium.

As demonstrated analytically in Refs.~\cite{wat15,wat20,hua19},
this kind of time crystal is in fact impossible, at least for all 
many-body systems with short-range interactions. Accordingly, 
also the above-mentioned phase transition can be ruled out.

An alternative, weaker definition of a time crystal has recently been
proposed and explored in Ref.~\cite{med20}, requiring that the ratio
between the temporal correlation in (\ref{57}) an its initial value
$C_{AB}(0)$ must exhibit permanent long-time oscillations.
According to (\ref{63}), this condition is always fulfilled for the
specific choice $A=B=M_x$.
In other words, according to this definition, a time crystal
is expected to generically arise for any model of the general 
form (\ref{1})-(\ref{3}) with non-vanishing field $h$.
Similarly to the discussion at the end of Sec.~\ref{s34},
our present findings thus complement
and substantially extend those obtained in 
the seminal previous Ref.~\cite{med20}.

\section{Summary and Conclusions}
\label{s7}

Our first main prediction, see Sec.~\ref{s3}, is that 
any Heisenberg model of the general 
form (\ref{1})-(\ref{3}) gives rise to 
time-dependent expectation values
(\ref{8a}), which become
practically indistinguishable from the 
auxiliary function (\ref{25}) for 
practically all sufficiently large times $t$.
The very weak
preconditions for this prediction are that 
the system size $\dof$ must be large, the maximal 
gap degeneracy $\gap$ must not be exceedingly large,
see below Eq.~\eqref{20}, 
and the maximal level population $\pmax$ must 
be small, see Eq.~(\ref{20}). 
For instance, the latter condition is known to be
fulfilled if the initial state arises as 
the result of a canonical quench, see Sec.~\ref{s33}.

In turn, this auxiliary function (\ref{25})
generically exhibits time-periodic 
but not necessarily harmonic oscillations,
hence the same must be (approximately) the case
for the long-time behavior of the corresponding
expectation values in (\ref{8a}),
as exemplified by Figs.~\ref{fig2}-\ref{fig5}.
The main requirements for such permanent  
long-time oscillations are a non-vanishing magnetic field 
$h$ in (\ref{2}), and a non-equilibrium initial 
condition (thus excluding diagonal 
ensembles of the form (\ref{26}) or (\ref{55})).
In particular, the system does not exhibit equilibration
in all these cases.

As detailed at the end of Sec.~\ref{s32},
the absence of equilibration can be
traced back to the existence of highly 
degenerate energy gaps, which in turn 
may be viewed as a consequence of the
spatially homogeneous external field and
the SU(2) symmetry of the field-free model.
On the other hand, whether or not the system
satisfies the so-called eigenstate thermalization 
hypothesis 
\cite{mor18,gog16,dal16}
does not seem to play a major role.

We remark that the considered models 
(\ref{1})-(\ref{3}) themselves are not
subject to any time-dependent external driving.
Moreover, all the above findings are
independent of whether the system is integrable or not,
features disorder and possibly many-body localization 
or not, is extensive due to short-range interactions or not,
nor does the dimensionality of the system play any 
significant role.

Put differently, approximately periodic long-time 
oscillations are predicted to occur for almost
any observable.
Moreover, during some initial time interval,
the expectation values are generically far
for from being periodic, and exhibit some small 
deviations from strict periodicity even for large times.
Finally, those oscillations are in general 
not of a purely harmonic character, including
as special cases oscillations with arbitrary 
multiples of the reference frequency $h$,
cf. Eq.~(\ref{25}).
As discussed in Sec.~\ref{s54}, for such 
observables we are thus unable to complement 
our analytical theory by some simple ``physical 
explanation'' of what is essentially going on.

Another challenging open problem is to explain 
all observable properties for a finite magnetic field 
$h$ in (\ref{2}) in terms of the field-free properties.
More precisely speaking, the eigenvalues and eigenvectors
are of course trivially related via (\ref{7}), (\ref{8}), 
but does the behavior of all physically relevant
observables for $h=0$ 
already determine their behavior for $h\not =0$\,?
For instance, the idea to switch into some suitable 
rotating frame might appear very natural and promising 
at first glance. However, closer inspection reveals
that for most observables the behavior for $h\not=0$ 
cannot be deduced from that for $h=0$ in this way.

Our second main result (see Sec.~\ref{s4})
is the prediction of synchronization under the 
additional requirement that the 
system is translationally invariant and thus obeys 
periodic boundary conditions in all spatial directions.
Here, the term synchronization means that the above discussed
long-time oscillations become approximately invariant under arbitrary 
translations of any given observable,
as exemplified by Figs.~\ref{fig4} and \ref{fig5}.
Once again, this approximate invariance is furthermore predicted 
to become asymptotically exact for large times and large 
system sizes.
Even more generally speaking, and without any reference 
to some underlying lattice geometry, it seems in fact
sufficient to require that all spins of the considered
model are equivalent,
and likewise for the synchronizing observables.

We emphasize that our present synchronization 
phenomenon does not depend
on whether the interactions $J_{ij}$ in (\ref{3}) 
are negative (i.e. of ferromagnetic character)
or not \cite{vor21}, contrary to what one might have naively 
expected to be necessary for the ``alignment'' of 
all the spins in such a system.
In the same vein, the system's dimensionality once again plays no role,
nor is it necessary that the initial condition is translationally
invariant.
More generally speaking, ordering and phase transition phenomena
at thermal equilibrium are apparently of little help to better 
understand our present synchronization effects,
nor are we able to provide any other kind of simple 
intuitive explanation of the basic underlying physics.

Obviously, 
the above predicted long-time oscillations
of any given observable $A$
in general still depend in a very complicated 
way -- via the phases and amplitudes in (\ref{25}) --
on the choice of the initial condition $\rho(0)$.
However, for translationally invariant 
Hamiltonians those long-time oscillations were 
shown in Sec. \ref{s4} to be invariant under 
arbitrary translations of the initial 
condition $\rho(0)$, even if $\rho(0)$ 
itself is not translationally invariant.
This quite remarkable finding is in fact 
equivalent to the prediction of synchronization,
and therefore seems again not to admit a 
simple physical explanation.

Our third main result concerns the issue of time crystals.
Unfortunately, even the precise definition of 
a time crystal still appears to be somewhat ambiguous.
For instance, already our permanent oscillations 
from Sec.~\ref{s3} can be considered as the characteristic
signature of a time crystal according to one of the definitions
provided in Ref.~\cite{ven19} (see Figure 8, second column,
last row therein):
Indeed, since the time-translation invariance of the model 
Hamiltonian is spontaneously broken and reduced to a 
time-discrete invariance for arbitrarily long times, 
which in turn may be viewed as a thermodynamic limit in the 
time domain, it seems justified~\cite{ven19} to
speak of a ``crystal'' in the time domain.
In our present explorations in Sec.~\ref{s6}, we mainly
focused on the somewhat more generally established definition
of a time crystal from Ref.~\cite{wat15}.
We also may recall that the no-go theorem for this type of 
time  crystals from Ref.~\cite{wat15} 
has been shown in Ref.~\cite{ven19}
to still contain a loophole, which in turn 
has been subsequently closed in \cite{wat20}, compare also \cite{hua19}.
Our present explorations are of course compatible
with this latter no-go theorem, i.e., we do not
find a time crystal in the sense of Ref.~\cite{wat15}.
Finally, yet another, somewhat weaker definition of a
time crystal has been proposed in Ref.~\cite{med20},
according to which our findings in Sec.~\ref{s6}
lead to the conclusion that models of the general
form (\ref{1})-(\ref{3}) generically do exhibit the
characteristic signature of a time crystal.
The question of what 
we actually gained by knowing whether or not
some given model system qualifies as a time crystal 
in one or the other 
sense remains unclear to the present authors.

Finally, it seems reasonable to expect that our 
main findings will also be recovered in a broad 
class of alternative models such as the Hubbard model,
as long as their general symmetry properties 
are similar as in our present model,
i.e., analogous the SU(2) symmetry 
of our field-free model and to the spatial 
homogeneity of the externally applied field.

\begin{acknowledgments}

This work was funded by the Deutsche Forschungsgemeinschaft 
(DFG, German Research Foundation) -- 
355031190 (FOR~2692), 397303734, and 397300368. 
We thank Heinz-J{\"u}rgen Schmidt for 
valuable suggestions and remarks.
We acknowledge support for the publication 
costs by the Open Access Publication Fund of Bielefeld University 
and the Deutsche Forschungsgemeinschaft (DFG).
\end{acknowledgments}

%
\appendix
\section{Derivation 
of Eq.~(\ref{19})}
\label{app1}

As usual, the unperturbed energies are denoted 
by $E_n^0$ with $n\in\{1,...,N\}$ (see below (\ref{6})),
and the operator norm (largest eigenvalue in modulus) 
of any Hermitian operator $A$ 
is denoted by $\norm{A}$.

Choosing $l = L_n$ in (\ref{5}), and exploiting that
$\norm{S^z}\leq \sum_{i\in \G }\norm{s_i^z}=\dof s$,
where $s$ is the 
single-site spin quantum number and $\dof$ the system size
(see above (\ref{1})),
we 
can conclude that
\begin{eqnarray}
L_n\leq \dof s
\label{a1}
\end{eqnarray}
for any $n\in\{1,...,N\}$.

Given that a single spin at any given site $i$ 
spans a Hilbert space of 
dimension $2s\! +\! 1 $, the dimensionality of the full Hilbert 
space will be $(2s\! +\! 1 )^{\dof}$.
Hence, 
the total number $N$ of all energy eigenvalues
$E_n^0$ can be {\em upper bounded} by $(2s\! +\! 1 )^{\dof}$,
\begin{eqnarray}
N\leq (2s + 1 )^{\dof}
\ .
\label{a2}
\end{eqnarray}
Conversely, for any given $n$, the total number 
$2L_n\! +\! 1$ of all possible labels $l$ 
(see below (\ref{6}))
is upper bounded by $2\dof s\! +\! 1 $ according to (\ref{a1}).
We thus obtain the {\em  lower bound}
\begin{eqnarray}
N\geq\frac{(2s + 1 )^{\dof}}{2\dof s + 1 }
\ .
\label{a3}
\end{eqnarray}
Altogether, (\ref{a2}) and (\ref{a3}) imply
that the number $N$ of energy eigenvalues 
$E_n^0$ must grow exponentially with the system size $\dof$

The set of all possible (ordered)
pairs of indices $m$ and $n$ is defined as
\begin{equation}
{\cal G}_{\rm tot}:=\bigl\{(m,n)\, |\, m ,n\in \{1,\dots,N\} \bigr\}
\ .
\label{a4}
\end{equation}
For any given pair $\alpha=(m,n)\in{\cal G}_{\rm tot}$
we furthermore define
\begin{eqnarray}
G_{\!\alpha} & := & E_n^0-E_m^0
\ ,
\label{a5}
\\
\eta_{\alpha}^\nu
& := & 
\sum_k  \rho_{mn}^{k,k+\nu} A_{nm}^{k+\nu,k}
\ .
\label{a6}
\end{eqnarray}
Hence, (\ref{13}) can be rewritten as
\begin{eqnarray}
f_{\nu}(t) & := & \sum_{\alpha\in{\cal G}_{\rm tot}} e^{iG_{\!\alpha} t} \, \eta_{\alpha}^\nu
\ .
\label{a7}
\end{eqnarray}

Next, we introduce the subset ${\cal G}\subset {\cal G}_{\rm tot}$ of all
pairs $(m,n)$ with the property that $E_m^0\not=E_n^0$,
i.e.,
\begin{equation}
{\cal G}  :=  \bigl\{\alpha \in {\cal G}_{\rm tot}\, |\, G_{\!\alpha}\not=0 \bigr\}
\ .
\label{a8}
\end{equation}
Accordingly, its complement satisfies
\begin{equation}
{\bar{\cal G}} :=  {\cal G}_{\rm tot}\!\setminus 
{\cal G} = \bigl\{\alpha \in {\cal G}_{\rm tot}\, |\, G_{\!\alpha}=0 \bigr\}
\ .
\label{a9}
\end{equation}

It readily follows that the maximal gap degeneracy from
(\ref{18}) can be rewritten in the form
\begin{eqnarray}
\gap & = & \max_{\beta\in{\cal G}} \left|\{ \alpha\in{\cal G} | \, G_\alpha=G_\beta\}\right|
\ ,
\label{a10}
\end{eqnarray}
where $|S|$ denotes the number of elements contained in the set $S$.
Similarly, the long-time average of $f_{\nu}(t)$ from (\ref{13}) or (\ref{a7})
can be rewritten in  the form (\ref{15}) or 
\begin{eqnarray}
\bar f_{\nu} = \sum_{\alpha\in {\bar{\cal G}}} \eta_\alpha^\nu
\ ,
\label{a11}
\end{eqnarray}
respectively.

As announced in the main text, our objective is 
to show  that the difference 
\begin{eqnarray}
\Delta(t) := \At - \Att
\label{a12}
\end{eqnarray}
between the time-dependent expectation values from
(\ref{12}) and the auxiliary function from (\ref{14})
is small for most sufficiently late times $t$.
Employing (\ref{12}), (\ref{14}), (\ref{a7}), (\ref{a9}), 
and (\ref{a11}), we therefore rewrite (\ref{a12}) as
\begin{eqnarray}
\Delta(t) & = & \sum_{\nu} \delta_{\nu}(t)\, e^{i\nu ht}
\ ,
\label{a13}
\\
\delta_{\nu}(t) & := & f_{\nu}(t)-\bar f_{\nu}=
\sum_{\alpha\in{\cal G}} e^{iG_{\!\alpha} t} \, \eta_{\alpha}^\nu
\ .
\label{a14}
\end{eqnarray}
Next we recall that the sum over the indices $m,n,k,l$ in (\ref{9}) 
is tacitly restricted to pairs $n,l$ for which $|n,l\rangle$ are 
well-defined eigenvectors in (\ref{7}), i.e. $n\in\{1,...,N\}$ and 
$l\in\{-L_n,...,L_n\}$,
and likewise for the pairs $m,k$.
Alternatively, for indices $n,l$ so that $|n,l\rangle$ is not a well-defined 
eigenvector, we may define those (so far undefined) vectors $|n,l\rangle$
as being equal to the null vector (hence $\rho_{mn}^{k,l}=0$, $A_{nm}^{l,k}=0$).
As a consequence, we may now consider all four indices
$m,n,k,l$ in the sum in (\ref{9}) to run over all integer values,
and likewise for the summation indices in (\ref{13}), (\ref{15}), 
and (\ref{a6}).
Furthermore, it follows that
the matrix elements $\rho_{mn}^{k,k+\nu}$ are zero if
$k\not\in\{-L_m,...,L_m\}$ or $k+\nu\not\in\{-L_n,...,L_n\}$.
Hence, it is sufficient to keep 
on the right-hand side in (\ref{a6}) 
only those 
summands
which satisfy $|k|\leq L_m$ 
and $|\nu+k|\leq L_n$.
Observing (\ref{a1}) and $|\nu+k|\geq |\nu|-|k|$
(triangle inequality) it follows that 
$|\nu|-\dof s\leq  |\nu|-|k| \leq |\nu+k|\leq \dof s$ 
must be fulfilled.
As a consequence, it is necessary that $|\nu|\leq 2\dof s$
in order that $\eta_\alpha^\nu$ in (\ref{a6})
is non-zero.
Therefore, it is sufficient to keep in (\ref{a13})
only those $\nu$ which are contained in 
$I:=\{-2\dof s,...,2\dof s\}$, and by employing the 
Cauchy-Schwarz inequality we obtain
\begin{eqnarray}
|\Delta(t)|^2 \leq \sum_{\nu\in I} |\delta_{\nu}(t)|^2
\sum_{\nu\in I} |e^{i\nu ht}|^2
\ .
\label{a15}
\end{eqnarray}
The last sum can be identified with $4\dof s\! +\! 1 $,
yielding
\begin{eqnarray}
|\Delta(t)|^2 & \leq & (4\dof s\! +\! 1 )\,  \sum_\nu |\delta_{\nu}(t)|^2
\ ,
\label{a16}
\end{eqnarray}
where, without loss of generality, the sum has 
again been extended to all integer indices $\nu$.

Denoting, as in the main text, 
the temporal average of an arbitrary function $f(t)$ over
the time interval $[0,T]$ by 
\begin{eqnarray}
\left\langle f(t)\right\rangle_{\! T}:=\frac{1}{T}\int_0^T \!\!dt\, f(t)
\ ,
\label{a17}
\end{eqnarray}
we can conclude from (\ref{a14}) that
\begin{eqnarray}
\left\langle |\delta_\nu(t)|^2 \right\rangle_{\! T}  & = &  \sum_{\alpha,\beta\in{\cal G}} (\eta^\nu_\alpha)^\ast
M^{\alpha\beta}_T \eta^\nu_{\beta}
\ ,
\label{a18}
\\
M^{\alpha\beta}_T & := & 
\left\langle e^{-i(G_\alpha-G_\beta)t}\right\rangle_{\! T} 
\ .
\label{a19}
\end{eqnarray}
Viewing $M^{\alpha\beta}_T$ as the matrix elements of some operator $M_T$,
one can infer from (\ref{a19}) that $M_T$ is Hermitian and non-negative,
and therefore
\begin{eqnarray}
\sum_{\alpha,\beta\in{\cal G}} (\eta^\nu_\alpha)^\ast M^{\alpha\beta}_T \eta^\nu_{\beta}
\leq
\norm{M_T}\sum_{\alpha\in{\cal G}} |\eta^\nu_\alpha|^2 
\ .
\label{a20}
\end{eqnarray}
As detailed, e.g., in Ref.~\cite{sho12}, compare Eq.~(14) therein, one can furthermore show that
\begin{eqnarray}
\norm{M_T} & \leq & 2\, \gap
\label{a21}
\end{eqnarray}
for all sufficiently large $T$, where $\gap$ is given in (\ref{a10}).

Altogether, (\ref{a16}), (\ref{a18}), (\ref{a20}), and (\ref{a21}) thus imply
\begin{eqnarray}
\left\langle |\Delta(t)|^2 \right\rangle_{\! T} & \leq &
2\, \gap
\, (4\dof s\! +\! 1 )\, \sigma^2
\ ,
\label{a22}
\\
\sigma^2 & := & \sum_\nu \sum_{\alpha\in{\cal G}} |\eta^\nu_\alpha|^2 
\ ,
\label{a23}
\end{eqnarray}
for all sufficiently large $T$.
Extending the sum in (\ref{a23}) over all index pairs $\alpha\in {\cal G}_{\rm tot}$
and exploiting (\ref{a6}), we find
\begin{eqnarray}
\sigma^2 & \leq &  \sum_\nu \sum_{mn} \sum_{kl} 
\rho_{mn}^{k,k+\nu} A_{nm}^{k+\nu,k}
(\rho_{mn}^{l,l+\nu} A_{nm}^{l+\nu,l})^\ast
\ ,
\nonumber
\\
& = & 
\sum_{\nu mn} Q_{\nu mn}
\ ,
\label{a24}
\\
Q_{\nu mn} & := & \sum_{kl} V_{\nu mn}^{k,l} (V_{\nu mn}^{l,k})^\ast 
\ ,
\label{a25}
\\
V_{\nu mn}^{k,l} & := & \rho_{mn}^{k,k+\nu} (A_{nm}^{l+\nu,l})^\ast
\ .
\label{a26}
\end{eqnarray}
Utilizing the Cauchy-Schwarz inequality in (\ref{a25}) implies
\begin{eqnarray}
|Q_{\nu mn}|^2\leq \sum_{kl} |V_{\nu mn}^{k,l}|^2 \sum_{kl} |V_{\nu mn}^{l,k}|^2
\ .
\label{a27}
\end{eqnarray}
Observing that the two sums on the right-hand side are in fact identical,
we can infer with (\ref{a24}) that
\begin{eqnarray}
\sigma^2 \leq \sum_{\nu mn} |Q_{\nu mn}| \leq \sum_{\nu mnkl} |V_{\nu mn}^{k,l}|^2 
\label{a28}
\end{eqnarray}
and with (\ref{a26}) that
\begin{eqnarray}
\sigma^2 \leq  \sum_{\nu mnkl} |\rho_{mn}^{k,k+\nu}|^2\, |A_{nm}^{l+\nu,l}|^2
\ .
\label{a29}
\end{eqnarray}

Exploiting (\ref{10}) and the Cauchy-Schwarz inequality,
one can conclude that 
$|\rho_{mn}^{k,l}|^2\leq \rho_{mm}^{k,k}\rho_{nn}^{l,l}$.
Since the density operator $\rho(0)$ must be semi-positive,
it follows with (\ref{10}) that $\rho_{mm}^{k,k}$ and 
$\rho_{nn}^{l,l}$ are non-negative, real numbers.
Altogether, $|\rho_{mn}^{k,k+\nu}|^2$ in (\ref{a29}) can thus be upper bounded by
$\rho_{mm}^{k,k}\pmax$, where $\pmax$ is defined in (\ref{17}),
yielding
\begin{eqnarray}
\sigma^2 & \leq & \pmax  \sum_{mk} \rho_{mm}^{k,k}\, W_{m}
\ ,
\label{a30}
\\
W_{m} & := & \sum_l w_{ml}
\ ,
\label{a31}
\\
w_{ml} & := & \sum_{n\nu} |A_{nm}^{l+\nu,l}|^2
\ .
\label{a32}
\end{eqnarray}
Replacing in (\ref{a32}) the summation index $\nu$ by $j:=l+\nu$ and exploiting
(\ref{11}) thus yields
\begin{eqnarray}
w_{ml}  =  \sum_{nj} |A_{nm}^{j,l}|^2 
=  \sum_{nj}\langle m,l | A | n,j \rangle \langle n,j |A| m,l \rangle
\, . \ \ \ \ \ \ \ 
\label{a33}
\end{eqnarray}
Since $\sum_{nj} | n,j\rangle \langle n,j|$ is the unit operator, we 
see that $w_{ml}$ equals $\langle m,l | A ^2 | m,l\rangle$
and thus 
\begin{eqnarray}
W_{m} = \sum_l \langle m,l | A ^2 | m,l\rangle
\ .
\label{a34}
\end{eqnarray}
As discussed below (\ref{a14}), the summands on the right-hand 
side of (\ref{a34})
are zero for $l\not\in\{-L_m,...,L_m\}$.
In other words, there are at most $2L_m\!+\!1$ 
non-vanishing summands. Furthermore, each of those
summands can be upper bounded by $\norm{A^2}=\norm{A}^2$.
Due to (\ref{a1}) we thus arrive at
\begin{eqnarray}
W_{m} \leq (2\dof s\! +\! 1) \,\norm{A}^2
\ .
\label{a35}
\end{eqnarray}
Observing (\ref{10}), the remaining sum in (\ref{a30}) can be 
identified with $\tr\{\rho(0)\}=1$, yielding
\begin{eqnarray}
\sigma^2 & \leq &   (2\dof s\! +\! 1 ) \,\norm{A}^2\,\pmax
\ .
\label{a36}
\end{eqnarray}
Together with (\ref{a22}), we finally can conclude that
\begin{eqnarray}
\left\langle |\Delta(t)|^2 \right\rangle_{\! T} & \leq &4\gap\, (2\dof s\! +\! 1 )^2\, \norm{A}^2\, \pmax
\label{a37}
\end{eqnarray}
for all sufficiently large $T$.

Note that  if we replace $A$ by $A+c$ then both terms on the right-hand side
in (\ref{a12}) are shifted by the same constant $c$, thus the
left-hand side is independent of $c$.
Accordingly, 
the left-hand side in (\ref{a37}) is independent of $c$, 
while the right-hand side yields in general a different
upper bound for different choices of $c$.
Denoting by $\amax$ and $\amin$ the largest and smallest eigenvalues of $A$,
respectively, one finds that the tightest upper bound is achieved 
for the choice $c=-(\amax+\amin)/2$.
Altogether, (\ref{a37}) and (\ref{a12}) thus yield 
\begin{eqnarray}
\left\langle [\At - \Att ]^2 \right\rangle_{T} 
& \leq &
\gap\, (2\dof s\! +\! 1 )^2\, \Da^2\, \pmax
\ \ \ 
\label{a38}
\end{eqnarray}
for all sufficiently large $T$,
where $\Da:=\amax-\amin$ is the measurement range of $A$
(difference between largest and smallest possible measurement 
outcomes).
In other words, we recover Eq.~(\ref{19}).

%
\section{Finite-Size Effects}
\label{app3}

The numerical examples considered in Sec.~\xref{s35} deal with 
still relatively
small spin systems. In particular, for the two-dimensional 
models
with open boundary conditions one might wonder 
how strongly finite-size effects impact the 
theoretically predicted
periodicity of time-dependent 
expectation values at late times. We also
remind the reader that synchronization is not to be expected 
in such models with open boundary conditions since
they are not translationally invariant.

\begin{figure}[ht!]
\includegraphics[scale=0.7]{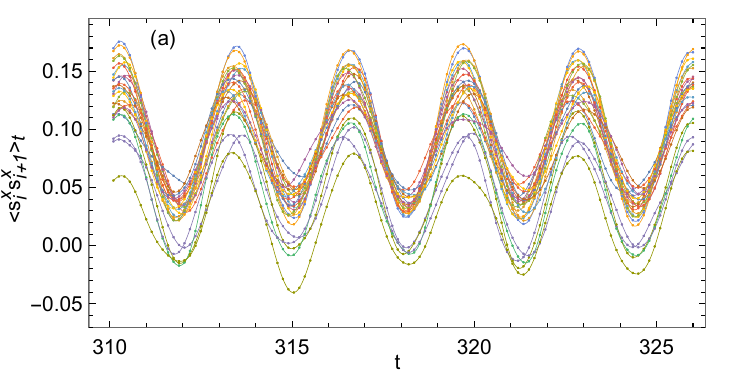}
\includegraphics[scale=0.7]{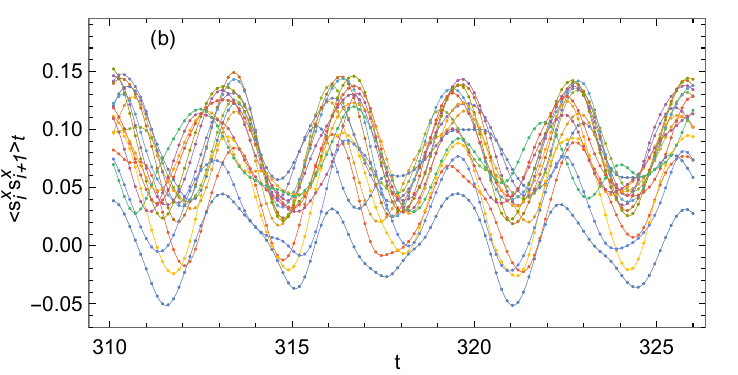}
\caption{
(a) 
Same numerical data as in Fig.~\ref{fig3}(d),
i.e., for a $5\times 5$ square lattice model 
with 
open boundary conditions.
(b): 
Corresponding data for a $4\times 4$ square lattice.
}
\label{fig6}
\end{figure}

In order to get an impression, \figref{fig6} compares the late
time behavior of spin-spin correlation functions for the already
shown example of a $5\times 5$ square lattice
(see \figref{fig3}(d))
with the smaller counterpart of a $4\times 4$ square lattice.
As is obvious for the naked eye,
the larger system displays much smaller deviations from
periodicity than the smaller system. 
Unfortunately, 
numerical explorations of even larger square lattices are 
prohibited
by the exponential growth of the underlying Hilbert 
space, but we think that already our present
comparison provides sufficient evidence that for 
larger systems better and better periodicity
is to be expected.

%
\section{Derivation of Eq.~(\ref{28})}
\label{app2}

We focus on spin models (\ref{2})
on a {\em one-dimensional} lattice $\G =\{1,...,\dof \}$ 
with { periodic boundary conditions}. 
[Generalizations to hypercubic lattices in arbitrary 
dimensions are straightforward.]
In other words, we are dealing with
$\dof$ { identical} ``units'' (spins) on a ring (chain with 
periodic boundary conditions) which are 
labeled by $i\in \{1,...,\dof \}$.

In the absence of interactions, each unit ``lives'' on 
a
Hilbert space $\hr_i$ with orthonormal basis $|k\rangle_i$, 
where $k=1,...,2s\!+\!1$.
Apart from ``belonging'' to different units $i$, all those
Hilbert spaces are identical copies of each other.

The pertinent Hilbert space $\hr$ of the total system is 
the tensor product of all the $\hr_i$.
Abbreviating $\dof$-tuples $(k_1,...,k_{\dof} )$ as
$\vec k$, the vectors
$|\vec k\rangle:=|k_1\rangle_1\cdots |k_{\dof} \rangle_{\dof} $
then amount to an orthonormal basis  of $\hr$.

Next, a ``shift'' or ``translation'' operator $\TT:\hr\to\hr$
is defined via its action on any basis vector: 
$\TT |\vec k\rangle :=|k_2\rangle_1|k_3\rangle_2\cdots |k_{\dof} \rangle_{\!\dof -1}|k_1\rangle_{\!\dof }$.
One readily concludes that $\TT$ is norm-preserving.
It follows that $\TT$ must be a unitary operator, i.e.,
 $\TT ^\dagger=\TT ^{-1}$.

Our main assumption is that the unperturbed Hamiltonian
$H_0$ from (\ref{3}) is {\em translationally invariant} in the
sense that the couplings $J_{ij}$ do not depend separately
on $i$ and $j$, but only on the 
difference $i-j$ (modulo $\dof$).
It follows that $H_0$ is also translationally invariant
in the alternative sense that 
\begin{eqnarray}
\TT^\dagger\! H_0\TT=H_0
\ ,
\label{b1}
\end{eqnarray}
or, equivalently, $[H_0,\TT]=0$ (commutator).
Likewise, one sees that each component of the
total spin $S^a$ from (\ref{1}) is translationally 
invariant.
It follows that all four operators 
$H_0$, $S^z$ , $\vec S^2$, and $\TT$
commute with each other. Without loss of generality,
we thus can 
assume that the eigenvectors $|n,l\rangle$
of $H_0$ are at the same time not only 
eigenvectors of $S_z$, and $\vec S^2$,
see (\ref{4})-(\ref{6}), but also eigenvectors of $\TT$.
Since $\TT$ is unitary, the corresponding eigenvalues
must be of unit modulus, i.e.,
\begin{eqnarray}
\TT|n,l\rangle = e^{i \theta_{n,l}}|n,l\rangle
\label{b2}
\end{eqnarray}
with certain ``phases'' $\theta_{n,l}\in[0,2\pi)$.

Since $S^x$ and $S^y$ commute with $\TT$
(see above),  the same applies to the raising operator
$S^+$ from (\ref{30}). Together with (\ref{31})
it follows that
\begin{eqnarray}
S^+\TT|n,l\rangle & = & e^{i \theta_{n,l}}S^+|n,l\rangle=e^{i \theta_{n,l}} c^+_{n,l} |n,l+1\rangle=
\nonumber
\\
\TT S^+|n,l\rangle & = &c^+_{n,l}  \TT  |n,l+1\rangle=c^+_{n,l}e^{i \theta_{n,l+1}}|n,l+1\rangle
\, . \ \ \ \ \ \
\label{b3}
\end{eqnarray}
We
thus can conclude that 
$\theta_{n,l+1}=\theta_{n,l}$, and finally that $\theta_{n,l}$  
only depends on $n$, but not on $l$.

Combining (\ref{b2}) with the $l$-independence of
$\theta_{n,l}$ one recovers (\ref{28})
for arbitrary Hermitian operators $B$.
As in the main text, it is {\em a priori} understood in 
(\ref{28}) that $n\in\{1,...,N\}$ and $l,l'\in \{-L_n,...,L_n\}$,
but with the convention adopted below (\ref{a14}),
one readily can extend the same relation 
to arbitrary 
$n,l,l'$.

We finally mention that the choice of the basis      
as specified below (\ref{b1}) may in principle not be 
unique, but that such ambiguities can be excluded 
if all energies $E_n^0$ are pairwise different,
as it is assumed at the beginning of Sec.~\ref{s4}.


\end{document}